\PassOptionsToPackage{unicode}{hyperref}
\PassOptionsToPackage{hyphens}{url}
\PassOptionsToPackage{dvipsnames,svgnames,x11names}{xcolor}
\documentclass[
  12pt]{article}

\usepackage{amsmath,amssymb}
\usepackage{graphicx}
\usepackage{booktabs}
\usepackage{caption}
\usepackage{array}
\usepackage{amsmath}       
\usepackage{iftex}
\usepackage{booktabs}
\usepackage{float}
\usepackage{adjustbox}
\usepackage{caption}

\ifPDFTeX
  \usepackage[T1]{fontenc}
  \usepackage[utf8]{inputenc}
  \usepackage{textcomp} % provide euro and other symbols
\else % if luatex or xetex
  \usepackage{unicode-math}
  \defaultfontfeatures{Scale=MatchLowercase}
  \defaultfontfeatures[\rmfamily]{Ligatures=TeX,Scale=1}
\fi
\usepackage{lmodern}
\ifPDFTeX\else
\fi
\IfFileExists{upquote.sty}{\usepackage{upquote}}{}
\IfFileExists{microtype.sty}{% use microtype if available
  \usepackage[]{microtype}
  \UseMicrotypeSet[protrusion]{basicmath} % disable protrusion for tt fonts
}{}
\makeatletter
\@ifundefined{KOMAClassName}{% if non-KOMA class
  \IfFileExists{parskip.sty}{%
    \usepackage{parskip}
  }{% else
    \setlength{\parindent}{0pt}
    \setlength{\parskip}{6pt plus 2pt minus 1pt}}
}{% if KOMA class
  \KOMAoptions{parskip=half}}
\makeatother
\usepackage{xcolor}
\makeatletter
\ifx\paragraph\undefined\else
  \let\oldparagraph\paragraph
  \renewcommand{\paragraph}{
    \@ifstar
      \xxxParagraphStar
      \xxxParagraphNoStar
  }
  \newcommand{\xxxParagraphStar}[1]{\oldparagraph*{#1}\mbox{}}
  \newcommand{\xxxParagraphNoStar}[1]{\oldparagraph{#1}\mbox{}}
\fi
\ifx\subparagraph\undefined\else
  \let\oldsubparagraph\subparagraph
  \renewcommand{\subparagraph}{
    \@ifstar
      \xxxSubParagraphStar
      \xxxSubParagraphNoStar
  }
  \newcommand{\xxxSubParagraphStar}[1]{\oldsubparagraph*{#1}\mbox{}}
  \newcommand{\xxxSubParagraphNoStar}[1]{\oldsubparagraph{#1}\mbox{}}
\fi
\makeatother

\usepackage{longtable,booktabs,array}
\usepackage{calc} % for calculating minipage widths
\usepackage{etoolbox}
\makeatletter
\patchcmd\longtable{\par}{\if@noskipsec\mbox{}\fi\par}{}{}
\makeatother
\IfFileExists{footnotehyper.sty}{\usepackage{footnotehyper}}{\usepackage{footnote}}
\makesavenoteenv{longtable}
\usepackage{graphicx}
\makeatletter
\def\maxwidth{\ifdim\Gin@nat@width>\linewidth\linewidth\else\Gin@nat@width\fi}
\def\maxheight{\ifdim\Gin@nat@height>\textheight\textheight\else\Gin@nat@height\fi}
\makeatother
\setkeys{Gin}{width=\maxwidth,height=\maxheight,keepaspectratio}
\makeatletter
\def\fps@figure{htbp}
\makeatother

\makeatletter
\@ifpackageloaded{caption}{}{\usepackage{caption}}
\AtBeginDocument{%
\ifdefined\contentsname
  \renewcommand*\contentsname{Table of contents}
\else
  \newcommand\contentsname{Table of contents}
\fi
\ifdefined\listfigurename
  \renewcommand*\listfigurename{List of Figures}
\else
  \newcommand\listfigurename{List of Figures}
\fi
\ifdefined\listtablename
  \renewcommand*\listtablename{List of Tables}
\else
  \newcommand\listtablename{List of Tables}
\fi
\ifdefined\figurename
  \renewcommand*\figurename{Figure}
\else
  \newcommand\figurename{Figure}
\fi
\ifdefined\tablename
  \renewcommand*\tablename{Table}
\else
  \newcommand\tablename{Table}
\fi
}
\@ifpackageloaded{float}{}{\usepackage{float}}
\floatstyle{ruled}
\@ifundefined{c@chapter}{\newfloat{codelisting}{h}{lop}}{\newfloat{codelisting}{h}{lop}[chapter]}
\floatname{codelisting}{Listing}

\makeatother
\makeatletter
\@ifpackageloaded{caption}{}{\usepackage{caption}}
\@ifpackageloaded{subcaption}{}{\usepackage{subcaption}}
\makeatother

\ifLuaTeX
  \usepackage{selnolig}  % disable illegal ligatures
\fi
\usepackage[]{natbib}
\usepackage{bookmark}

\IfFileExists{xurl.sty}{\usepackage{xurl}}{} % add URL line breaks if available
\graphicspath{{./images/}{./plots/}}

\hypersetup{
  pdftitle={Using Bayesian Sequential Logistic Regression to Model the Organ Transplant Selection Process},
  pdfauthor={Xukun Zhu; Tyler Schappe; Ursula Rogers; Roland Matsouaka; Lisa M. McElroy; Samuel I. Berchuck},
  pdfkeywords={continuation-ratio model, partial pooling, structural attrition, multi-stage clinical continuum, social determinants of health},
  colorlinks=true,
  linkcolor={blue},
  filecolor={Maroon},
  citecolor={Blue},
  urlcolor={Blue},
  pdfcreator={LaTeX via pandoc}}

\newcommand{\anon}{1}

\begin{document}

\def\spacingset#1{\renewcommand{\baselinestretch}%
{#1}\small\normalsize} \spacingset{1}

%%%%%%%%%%%%%%%%%%%%%%%%%%%%%%%%%%%%%%%%%%%%%%%%%%%%%%%%%%%%%%%%%%%%%%%%%%%%%%

\if1\anon
{
  \title{\bf Using Bayesian Sequential Logistic Regression to Model the Organ Transplant Selection Process}
  \author{
    Xukun Zhu\(^{1}\), Tyler Schappe\(^{2}\), Ursula Rogers\(^{2}\),
    Roland Matsouaka\(^{3}\),\\
    Jacqueline B. Henson\(^{4}\), Norine W. Chan\(^{2}\),
    Goni Katz-Greenberg\(^{4}\),\\
    Lisa M. McElroy\(^{2,5}\), and Samuel I. Berchuck\(^{1,3}\)\\[2ex]
    \(^{1}\)Department of Statistical Science, Duke University\\
    \(^{2}\)Department of Surgery, Duke University School of Medicine\\
    \(^{3}\)Department of Biostatistics and Bioinformatics,\\
    \hspace*{1em}Duke University School of Medicine\\
    \(^{4}\)Department of Medicine, Duke University School of Medicine\\
    \(^{5}\)Department of Population Health Sciences,\\
    \hspace*{1em}Duke University School of Medicine}
  \maketitle
} \fi

\if0\anon
{
  \bigskip
  \bigskip
  \bigskip
  \begin{center}
    {\LARGE\bf Using Bayesian Sequential Logistic Regression to Model the
    Organ Selection Process}
\end{center}
  \medskip
} \fi

\bigskip
\begin{abstract}
Organ transplantation evaluation is a multi-stage process in which only patients who complete one stage remain eligible for the next, leading to patient attrition and reducing later-stage sample sizes. Covariate associations may differ across decision points, and identifying these stage-specific effects is necessary to target interventions and understand disparities. Conventional approaches that model each transition independently fail to distinguish stage-specific from shared associations and tend to produce imprecise estimates at later stages. We propose a Bayesian sequential logistic regression (BSLR) framework that represents progression as conditional Bernoulli distributions within a joint likelihood that preserves chronological ordering; we further show its equivalence to an adjusted continuation-ratio model. A hierarchical prior decomposes each covariate effect into an overall effect and a transition-specific deviation. This partial-pooling structure shares information across stages and stabilizes estimation where data are sparse, while a masking mechanism accommodates covariates structurally unavailable at particular stages without discarding records or imputing values. In simulations, BSLR achieved lower bias, posterior variance, and mean squared error than independent logistic regression. We demonstrated the utility of our model using a cohort of patients referred for kidney or liver transplant. Posterior estimates indicated that the associations of clinical indicators and social determinants of health with progression differed across care continuum phases. BSLR identified stage-specific heterogeneity in covariate associations, revealing factors whose effects reversed direction across evaluation stages and detecting associations at late-stage transitions that independent models failed to resolve due to insufficient precision.
\end{abstract}

\noindent%
\vfill

\newpage
\spacingset{1.8} % DON'T change the spacing!

\section{Introduction}\label{sec1}

\begin{figure}[tbp]
    \centering
    \includegraphics[width=\textwidth]{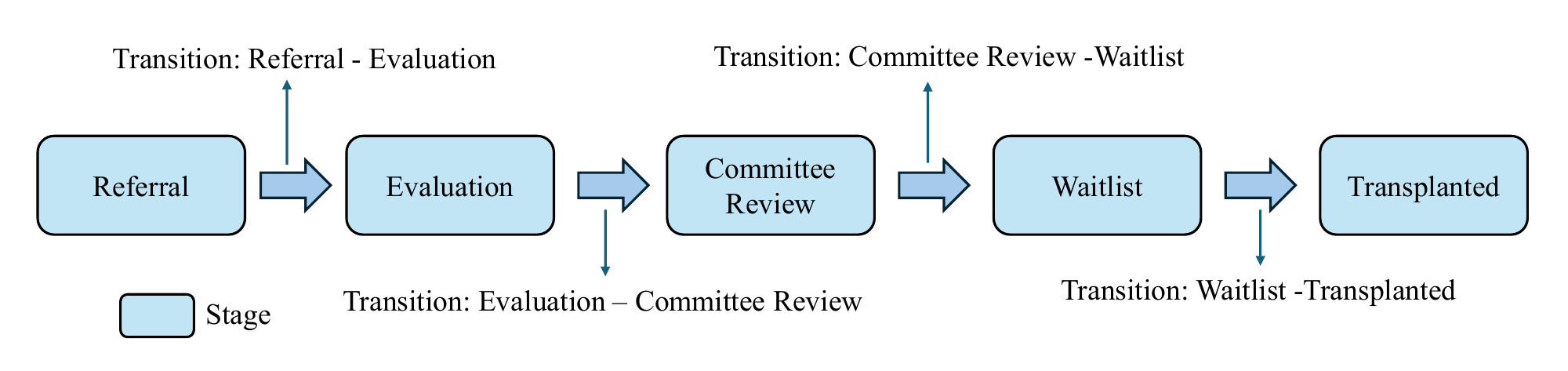}
    \caption{The multi-stage organ transplant evaluation process. Patient
    progression is tracked across four sequential transitions: Referral to
    Evaluation (Ref-Eval), Evaluation to Committee Review (Eval-CR),
    Committee Review to Waitlist (CR-WL), and Waitlist to Transplant
    (WL-TX).}
    \label{fig:flowchart}
\end{figure}

Organ transplantation is the standard treatment for patients with end-stage organ failure \citep{bezinover2019organ}. The patient selection for transplantation is a multi-stage sequential process from referral to transplanted \citep{watson2012organ}. At each stage, patient attrition may be influenced by clinical factors, such as disease severity and physiological comorbidities, as well as social determinants of health, including health literacy, insurance coverage, and geographic distance to transplant centers \citep{DeutschLink2021, Berchuck2024Predictive, 10.1001/jamanetworkopen.2025.49629}. Consequently, analyzing the factors associated with patient attrition across this continuum is necessary for improving the efficiency of the evaluation process and ensuring equity in organ allocation. Figure~\ref{fig:flowchart} illustrates the process a patient goes through during an organ transplant. Recent studies have developed predictive models to determine transplant necessity and evaluate factors influencing candidates' chances of receiving an organ. Bertsimas et al.\ used classification trees to address limitations of the Model for End-Stage Liver Disease (MELD) score, predicting three-month waitlist mortality or removal for liver transplant candidates \citep{bertsimas2019development}. Ascher-Bartlett et al.\ developed a logistic regression model to predict spontaneous native liver survival versus the need for transplantation or death in pediatric acute liver failure. Utilizing LASSO regression for feature selection, their model identified clinical predictors such as albumin, ammonia, and total bilirubin, with predictive performance comparable to the pediatric end-stage liver disease score \citep{ascher2024chalf}. In the context of heart transplantation, Zhang et al.\ developed the US Candidate Risk Score. By incorporating time-varying clinical, laboratory, and hemodynamic data, their model yielded higher discrimination for predicting short-term waitlist mortality compared to existing risk stratification tools \citep{zhang2024development}. Despite these developments, existing research focuses on predicting cross-sectional endpoints rather than evaluating factors that influence patient attrition at each stage of the transplantation continuum. Furthermore, these predictive frameworks restrict their scope to clinical variables, omitting the impact of non-clinical factors across the evaluation process.

To address the nature of multi-stage processes, sequential logistic regression is often utilized. Sequential logistic regression models stage-based outcomes by decomposing them into a series of binary transitions and has been applied across various research fields, including epidemiology, health services research and social science \citep{Lee2010, rigon2019bayesian, wideman2009prospective, dissanayake2002analysis, deppen2014predicting}. However, classical implementations estimate the parameters of each transition independently, which introduces two limitations when applied to the organ transplantation process. First, because each transition is fitted in isolation, these models evaluate the impact of a variable only on a single decision rather than assessing its overall effect across the entire sequential continuum \citep{powers2008statistical, agresti2010analysis}. Second, structural patient attrition reduces the sample entering each subsequent stage, so that the later transitions are estimated from few events; independent estimation therefore yields imprecise and unstable parameter estimates at later stages. These issues pose methodological challenges for accurately evaluating how different factors influence patient progression throughout the complete organ transplant care continuum.

Motivated by these challenges, we propose a Bayesian sequential logistic regression (BSLR) framework to model patient attrition in the organ transplantation evaluation process. The BSLR framework formulates the multi-stage progression as a sequence of conditional Bernoulli distributions within a joint likelihood that respects the chronological ordering of transitions, and we show that this formulation is equivalent to an adjusted continuation-ratio model. To address the limitations of independent transition models, the framework incorporates a hierarchical prior structure that decomposes covariate effects into an overall effect and a transition-specific deviation \citep{hoff2009first, seltzer1996bayesian}. This partial pooling mechanism facilitates information sharing across the sequential stages, stabilizing parameter estimates for late-stage transitions where sample sizes are reduced. In addition, a masking mechanism accommodates covariates that are absent at particular stages. Consequently, this framework enables the simultaneous estimation of overall and stage-specific effects of clinical variables and social determinants of health across the care continuum.

The remainder of this paper is organized as follows. Section~\ref{sec2} details the proposed BSLR framework. Section~\ref{sec3} presents a simulation study to compare the performance of the BSLR model with standard independent logistic regression (ILR). In Section~\ref{sec4}, we apply the framework to an organ transplantation dataset to analyze the factors associated with patient attrition across the evaluation continuum. Finally, Section~\ref{sec5} concludes the paper with a discussion of the findings.

\section{Methods}\label{sec2}

\subsection{Notation}
\label{subsec:notation}

To formalize the multi-stage organ transplant process, we first define the notations used throughout this article. Consider a stage-dependent process observed for $N$ independent patients, consisting of $S$ sequential stages, indexed by $s \in \{1, \dots, S\}$, and $J = S - 1$ transitions, indexed by $j \in \{1, \dots, J\}$, where transition $j$ refers to the attempted move from stage $j$ to stage $j+1$.

\begin{itemize}
    \item $Y_i \in \{1, 2, \dots, S\}$ denotes the categorical random variable representing the final evaluation stage reached by patient $i$ ($i = 1, \dots, N$). The event $Y_i = s$ indicates that the patient successfully reached stage $s$ but failed to advance further, so that $Y_i = S$ indicates the completion of all $J$ transitions. We write $\phi_{is} = P(Y_i = s)$ for the corresponding cell probabilities.
    \item $W_{ij} \in \{0,1\}$ is a binary indicator of a successful transition, linked to the categorical endpoint by the deterministic relationship
    \begin{equation}
        W_{ij} = \mathbb{I}(Y_i \geq j+1), \qquad j = 0, 1, \dots, J,
        \label{eq:deterministic}
    \end{equation}
   so that, for $j\ge 1$, $W_{ij}=1$ if patient $i$ moves from stage $j$ to stage $j+1$ and $0$ otherwise. Definition~\eqref{eq:deterministic} is stated for $j = 0, \dots, J$ so that $W_{i0} = \mathbb{I}(Y_i \geq 1) \equiv 1$ holds automatically rather than by convention; only $j = 1, \dots, J$ index modelled transitions. Definition~\eqref{eq:deterministic} determines $W_{ij}$ for every $j$ once $Y_i$ is known. For a patient who did not reach stage $j$, that is $Y_i < j$, the indicator is degenerate rather than missing: $W_{i,j-1} = 0$ forces $W_{ij} = 0$ with probability one, and the corresponding likelihood factor equals one (Appendix~A, Lemma~A.1). Transition $j$ is therefore informative only for the patients at risk, those with $Y_i \geq j$.
    \item $X_{i,j} \in \mathbb{R}^{P_{tv}}$ represents the column vector of time-varying covariates for patient $i$ measured at stage $j$, prior to the attempted transition $j$.
    \item $Z_i \in \mathbb{R}^{P_{fix}}$ represents the column vector of time-invariant covariates for patient $i$, excluding the intercept.
    \item $M_j \in \{0, 1\}^{P_{tv}}$ is a binary mask column vector used to explicitly address the structural availability of specific time-varying clinical measurements at transition $j$. We write $\mathcal{J}_p = \{\, j : M_{p,j} = 1 \,\}$ for the set of transitions at which time-varying covariate $p$ is available.
    \item $\{\alpha_j\}_{j=1}^{J}$ with $\alpha_j \in \mathbb{R}$, $\{\beta_j\}_{j=1}^{J}$ with $\beta_j \in \mathbb{R}^{P_{tv}}$, and $\{\gamma_j\}_{j=1}^{J}$ with $\gamma_j \in \mathbb{R}^{P_{fix}}$ represent the transition-specific intercepts, time-varying covariate effects, and time-invariant covariate effects, respectively.
    \item $\alpha \in \mathbb{R}, \beta \in \mathbb{R}^{P_{tv}}, \gamma \in \mathbb{R}^{P_{fix}}$ denote the overall, transition-agnostic baseline effects for the intercept, time-varying covariates, and time-invariant covariates, respectively.
    \item $\{a_j\}_{j=1}^{J}$ with $a_j \in \mathbb{R}$, $\{b_j\}_{j=1}^{J}$ with $b_j \in \mathbb{R}^{P_{tv}}$, and $\{g_j\}_{j=1}^{J}$ with $g_j \in \mathbb{R}^{P_{fix}}$ denote the transition-specific deviations from the overall baseline effects at transition $j$.
    \item $\theta$ denotes the full parameter vector, collecting the overall effects and the transition-specific deviations introduced in Section~\ref{subsec:model}; all probabilities below are conditional on $\theta$, on $Z_i$ and on a fixed covariate path $X_{i,1:J}=(X_{i,1},\dots,X_{i,J})$, and this conditioning is suppressed where it is not needed. Under the specification of Section~2.3, the likelihood depends on this path
only through the components at stages actually reached (see~\eqref{eq:cat_lik_truncated}), so covariates at stages a patient never entered need not be observed or defined.
\end{itemize}
Throughout, the at-risk event for transition $j$ is \begin{equation*} \{Y_i \geq j\} \;=\; \{W_{i,j-1} = 1\} \;=\; \{\text{patient } i \text{ reached stage } j\}, \end{equation*} so that ``occupying stage $j$'', ``being at risk for transition $j$'' and, for $j\ge 2$, ``having completed transition $j-1$'' describe the same event. Each transition is labelled by its stage of origin, so the transition out of stage $j$ is transition $j$. We use $s$ for quantities indexed by stage and $j$ for quantities indexed by transition; an event such as $\{Y_i\ge j\}$ compares the final stage with the origin stage of transition $j$, and $\{Y_i\ge j+1\}$ means that transition $j$ succeeded.

\subsection{Likelihood} \label{subsec:likelihood} 
The organ transplant selection process is a multi-stage continuum. In this clinical setting, the continuum is defined by five patient stages ($S = 5$), yielding four sequential transitions ($J = 4$): (1) Referral to Evaluation, (2) Evaluation to Committee Review, (3) Committee Review to Waitlist, and (4) Waitlist to Transplant. This process exhibits a chronological dependence: to be evaluated for transition $j\ge 2$, a patient must first complete
transition $j-1$.

Standard statistical approaches model such sequential categorical outcomes using continuation-ratio (CR) logit models \citep{agresti2010analysis}, which are conventionally parameterized through the conditional probability that a patient stops at the stage currently occupied. Because every quantity reported in this paper is an effect on progression along the continuum, we work throughout with the complementary conditional probability of advancing,
\begin{equation}
\omega_{ij} = P(Y_i \ge j+1 \mid Y_i \ge j), \qquad j = 1, \ldots, J,
\label{eq:adjusted_cr}
\end{equation}
so that a coefficient above zero on the logit scale corresponds to a higher chance of moving to the next stage. This is the standard forward parameterization of the sequential class of ordinal models \citep{tutz1991sequential}.

In practice, multi-stage clinical continuums are most naturally modelled through a sequence of transition-specific binary outcomes $W_{ij}$, indicating whether patient $i$ successfully moves from stage $j$ to stage $j+1$. The corresponding transition probability is the conditional probability of a successful transition given that the patient had reached stage $j$:
\begin{equation}
\pi_{ij} = P(W_{ij}=1 \mid W_{i,j-1}=1), \qquad j = 1, \ldots, J,
\label{eq:transition_prob}
\end{equation}
which is well defined at $j = 1$ because $W_{i0} \equiv 1$ by~\eqref{eq:deterministic}. In what follows we establish that this Bernoulli formulation is mathematically equivalent to the categorical formulation based on~\eqref{eq:adjusted_cr}, first at the level of transition probabilities and then at the level of the full sample likelihood.

We begin with the transition-probability level. Relationship \eqref{eq:deterministic} gives $$ P(W_{ij}=1 \mid Y_i = s) = \begin{cases} 1, & s \geq j+1, \\ 0, & s \leq j, \end{cases} \qquad s \in \{1, \ldots, S\}. $$ Since $\{W_{ij}=1\}=\{Y_i\ge j+1\}\subseteq\{Y_i\ge j\}=\{W_{i,j-1}=1\}$, the definition of conditional probability gives \[ \pi_{ij}=\frac{P(W_{ij}=1,\,W_{i,j-1}=1)}{P(W_{i,j-1}=1)} =\frac{P(W_{ij}=1)}{P(W_{i,j-1}=1)}, \] so that $\pi_{ij}$ is a ratio of marginal probabilities. Applying the law of total probability with respect to $Y_i$ to its numerator and denominator, the binary transition probability reduces to the continuation ratio,
\begin{equation}
\pi_{ij} = \frac{P(W_{ij}=1)}{P(W_{i,j-1}=1)} = \frac{P(Y_i \geq j+1)}{P(Y_i \geq j)} = \omega_{ij};
\label{eq:equivalence}
\end{equation}
the intermediate steps are given in Appendix~A. Thus $\pi_{ij}$ coincides exactly with $\omega_{ij}$, so that any logit-linear regression model posited for $\omega_{ij}$ (introduced in Section~\ref{subsec:model}) applies verbatim to $\pi_{ij}$.

We now turn to the likelihood level. To make the parallel between the two formulations transparent, we derive the sample likelihood independently under each formulation and then verify that the resulting expressions coincide.

Under the categorical formulation $Y_i \sim \text{Categorical}(\phi_{i1}, \ldots, \phi_{iS})$, the contribution of patient $i$ is the cell probability of the realized stage,
\begin{equation}
L_i^{(C)}(\theta) = P(Y_i = y_i \mid \theta) = \phi_{i, y_i}.
\label{eq:cat_lik_raw}
\end{equation}
To re-express $\phi_{i, y_i}$ in terms of the continuation ratios, iterate $\omega_{ij} = P(Y_i \geq j+1 \mid Y_i \geq j)$ together with $P(Y_i \geq 1) = 1$ to obtain the cumulative recursion
\begin{equation}
P(Y_i \geq s) = \prod_{l=1}^{s-1} \omega_{il}, \qquad s = 1, \ldots, S,
\label{eq:cumulative}
\end{equation}
where the product is taken over transitions $l$ and an empty product equals one. Combined with the difference identity $\phi_{is} = P(Y_i \geq s) - P(Y_i \geq s+1)$, this yields the continuation-ratio factorization of the cell probabilities,
\begin{equation}
\phi_{is} = \Bigg(\prod_{l=1}^{s-1} \omega_{il}\Bigg)(1 - \omega_{is}) \;\; \text{for } s < S, \qquad \phi_{iS} = \prod_{l=1}^{J} \omega_{il}.
\label{eq:cr_factorization}
\end{equation}
By the labelling convention of Section~2.1, $\omega_{is}$ is the probability of transition $s$, which exists because $s<S$ implies $s\le J$. Using the indicators of~\eqref{eq:deterministic}, evaluated at the observed
$y_i$, the cell probabilities~\eqref{eq:cr_factorization} can be assembled into the unified representation
\begin{equation}
L_i^{(C)}(\theta) \;=\; \phi_{i, y_i} \;=\; \prod_{j=1}^{J} \left[ \omega_{ij}^{W_{ij}} (1 - \omega_{ij})^{1 - W_{ij}} \right]^{W_{i,j-1}}.
\label{eq:cat_lik_compact}
\end{equation}
The case-by-case verification of this identity, distinguishing $y_i < S$ from $y_i = S$, is deferred to Appendix~A. Intuitively, the outer exponent $W_{i,j-1} = \mathbb{I}(Y_i \geq j)$ acts as an observation switch: the $j$-th factor enters the likelihood only when the patient was actually at risk for the $j$-th transition.

Equivalently, since the outer exponent vanishes for $j > y_i$, the product in~\eqref{eq:cat_lik_compact} truncates to
\begin{equation}
L_i^{(C)}(\theta) \;=\; \prod_{j=1}^{\min(y_i,\,J)}
\omega_{ij}^{W_{ij}} (1 - \omega_{ij})^{1 - W_{ij}},
\label{eq:cat_lik_truncated}
\end{equation} 
which makes explicit that the contribution of patient $i$ involves only the covariates $X_{i,1}, \dots, X_{i,\min(y_i,\,J)}$ recorded at the stages actually reached. This is consistent with the imputation strategy of Section~4, where stage-$j$ covariates are imputed only for patients who reached stage $j$.

We next derive the corresponding expression under the Bernoulli formulation, which models each observed transition as a conditional Bernoulli draw,
\begin{equation}
W_{ij} \mid \{W_{i,j-1} = 1\} \sim \text{Bernoulli}(\pi_{ij}), \quad j = 1, \ldots, J.
\label{eq:bernoulli_model}
\end{equation}
By the chain rule, the joint probability of $(W_{i1},\dots,W_{iJ})$ factorizes into the conditional probabilities of each $W_{ij}$ given its predecessors. Because the sequence is non-increasing, the $j$-th conditional is the Bernoulli$(\pi_{ij})$ probability when $W_{i,j-1}=1$, and equals one when $W_{i,j-1}=0$, since $W_{ij}=0$ then holds with probability one (Appendix~A, Lemma~A.1). Using $W_{i,j-1}$ as an exponent captures both cases and yields the closed-form contribution
\begin{equation}
L_i^{(B)}(\theta) = \prod_{j=1}^{J} \left[ \pi_{ij}^{W_{ij}} (1 - \pi_{ij})^{1 - W_{ij}} \right]^{W_{i,j-1}}.
\label{eq:bernoulli_likelihood}
\end{equation}

Expressions~\eqref{eq:cat_lik_compact} and~\eqref{eq:bernoulli_likelihood} now share an identical algebraic structure, and the transition-probability equivalence~\eqref{eq:equivalence} gives $\pi_{ij} = \omega_{ij}$, so that $L_i^{(C)}(\theta) = L_i^{(B)}(\theta)$ for every patient $i$. Independence across patients then gives the sample-level identity
\begin{equation}
L^{(C)}(\theta)
\;=\; \prod_{i=1}^{N} \phi_{i, y_i}
\;=\; \prod_{i=1}^{N} \prod_{j=1}^{J} \left[ \pi_{ij}^{W_{ij}} (1 - \pi_{ij})^{1 - W_{ij}} \right]^{W_{i,j-1}}
\;=\; L^{(B)}(\theta).
\label{eq:likelihood_equivalence}
\end{equation}
At the level of a single patient, the identity $L_i^{(C)}(\theta)=L_i^{(B)}(\theta)$ follows from the deterministic relationship~\eqref{eq:deterministic} and the law of total probability alone, and imposes no modelling assumption. The sample-level identity~\eqref{eq:likelihood_equivalence} additionally requires that patients be conditionally independent given their covariates and $\theta$, an assumption shared by both formulations. All remaining modelling assumptions therefore enter through the regression specification of Section~2.3, in which $\pi_{ij}$ depends on the patient's history only through $X_{i,j}$ and $Z_i$ and not on the time already spent in stage $j$; robustness to this assumption is assessed in Section~3.3.

The likelihood equivalence~\eqref{eq:likelihood_equivalence} guarantees that posterior inference on $\theta$ is invariant to the choice between the categorical and Bernoulli formulations, so that the same fit admits both a transition-level and a categorical-outcome interpretation. We adopt the Bernoulli form~\eqref{eq:bernoulli_likelihood} in the remainder of the paper because (i) the indicator $W_{i,j-1}$ naturally encodes the observation structure of sequential clinical continuums, ensuring that each patient contributes only to the transitions for which they were actually at risk, and (ii) it is substantially more convenient to express and sample from in modern probabilistic programming environments, while the categorical form~\eqref{eq:cr_factorization} remains available in closed form for reporting stage probabilities along a specified covariate path $X_{i,1:J}$.

\subsection{Model specification}
\label{subsec:model}

Having established the Bernoulli likelihood, we specify a sequential logit model for the transition probabilities $\pi_{ij}$ to evaluate covariate effects across the sequence:
\begin{equation}
\text{logit}(\pi_{ij}) = \alpha_j + X_{i,j}^{\top}\beta_j + Z_i^{\top} \gamma_j,
\qquad j = 1, \dots, J.
\label{eq:categorical_logit}
\end{equation}
Here, $X_{i,j}$ and $Z_i$ are the column vectors for time-varying and time-invariant covariates, respectively. The parameters $\alpha_j$, $\beta_j$, and $\gamma_j$ represent the transition-specific intercept and covariate effects. To address estimation instability caused by sample depletion at later stages, we utilize a hierarchical prior structure that decomposes these regression parameters into an overall shared effect and a transition-specific deviation:
\begin{equation}
\alpha_j = \alpha + a_j, \quad \beta_j = \beta + b_j, \quad \gamma_j = \gamma + g_j,
\qquad j = 1, \dots, J.
\label{eq:decomposition}
\end{equation}

In this hierarchical structure, $\alpha$, $\beta$, and $\gamma$ capture the overall baseline trends shared across the entire evaluation continuum. The terms $a_j$, $b_j$, and $g_j$ capture the unique transition-specific shifts from this baseline at transition $j$. The hierarchical prior distributions for these components are defined as follows:
\begin{align}
\alpha &\sim \mathcal{N}(0, \sigma_{\alpha}^2), \quad &a_j &\sim \mathcal{N}(0, \sigma_{a}^2) \nonumber \\
\beta_p &\sim \mathcal{N}(0, \sigma_{\beta,p}^2), \quad &b_{p,j} &\sim \mathcal{N}(0, \sigma_{b,p}^2), \quad \text{for } p=1, \dots, P_{tv} \nonumber \\
\gamma_q &\sim \mathcal{N}(0, \sigma_{\gamma,q}^2), \quad &g_{q,j} &\sim \mathcal{N}(0, \sigma_{g,q}^2), \quad \text{for } q=1, \dots, P_{fix}
\label{eq:priors}
\end{align}
By separating the variance hyperparameters for the overall effects ($\sigma_{\alpha}, \sigma_{\beta}, \sigma_{\gamma}$) and the deviations ($\sigma_{a}, \sigma_{b}, \sigma_{g}$), this partial pooling mechanism allows the baseline magnitude of a covariate to differ flexibly from its propensity to fluctuate across stages. Since the likelihood depends on these parameters only through the sums in~\eqref{eq:decomposition}, the overall effects and the deviations are not separately identified, and we impose the sum-to-zero constraints $\sum_{j=1}^{J} a_j = 0$, $\sum_{j \in \mathcal{J}_p} b_{p,j} = 0$ and $\sum_{j=1}^{J} g_{q,j} = 0$, where $\mathcal{J}_p$ is as defined in Section~\ref{subsec:notation}; the priors in~\eqref{eq:priors} are understood as conditioned on these constraints, which are implemented in Stan through a sum-to-zero parameterization.

However, real-world sequential clinical data inherently contain structural missingness. For example, the estimated glomerular filtration rate (eGFR) score is unmeasured during early administrative stages but serves as a metric during late-stage allocation. To mathematically accommodate this structural unavailability within our binary framework, we introduce the binary mask column vector $M_j \in \{0, 1\}^{P_{tv}}$. The $p$-th element $M_{p,j} = 1$ if the $p$-th time-varying covariate is clinically evaluated at stage $j$, and $M_{p,j} = 0$ if it is structurally unmeasured. Integrating $M_j$ directly into the linear predictor via the Hadamard (element-wise) product ($\odot$) formalizes the masking mechanism and yields the final regression model:
\begin{equation}
\text{logit}(\pi_{ij}) = \alpha_j + X_{i,j}^{\top}(M_j \odot \beta_j) + Z_i^{\top} \gamma_j,
\qquad j = 1, \dots, J.
\label{eq:masked_linear_predictor}
\end{equation}
When $M_{p,j} = 0$, the term $X_{i,j,p}\beta_{p,j}$ is masked from the linear predictor. This design gracefully removes the unmeasured covariate's impact on the joint likelihood for transition $j$ without discarding patient records or relying on arbitrary data imputation strategies.

\subsection{Prior and Computational Details}
\label{subsec:computation}
The hierarchical specification is completed by assigning priors to the standard-deviation hyperparameters governing the overall effects and the transition-specific deviations. Following the recommendations of \citet{gelman2006prior} for variance parameters in hierarchical models, we place weakly informative Half-Student-$t$ priors with three degrees of freedom and scale $3$ on all scale parameters $\sigma_{\alpha}, \sigma_{a}, \sigma_{\beta,p}, \sigma_{b,p}, \sigma_{\gamma,q}$ and $\sigma_{g,q}$. The half-$t$ family is supported on $[0, \infty)$ with its mode at the origin, so no location parameter is required. These priors place their mode at zero, thereby regularizing the deviation terms and encouraging pooling across transitions, while their heavy tails still permit genuinely large transition-specific effects to emerge when supported by the data \citep{gelman2015stan}. As a sensitivity check, we refit the model with one degree of freedom, corresponding to the Half-Cauchy$(3)$ prior advocated by \citet{gelman2006prior} and \citet{polson2012half}, which places still greater mass in the tails.

We implement the model in Stan \citep{carpenter2017stan, gelman2015stan}, which performs full Bayesian inference using Hamiltonian Monte Carlo through its No-U-Turn Sampler \citep{hoffman2014nuts}, an adaptive variant that automatically tunes trajectory length and step size during warm-up. We run four parallel chains of $100{,}000$ iterations each, discarding the first $50{,}000$ as warm-up, which yields $200{,}000$ post-warm-up draws for posterior summaries. Convergence is assessed using the rank-normalized potential scale reduction factor $\widehat{R}$ and the effective sample size \citep{vehtari2021rank}.

Full source code, including the Stan model implementation and reproducible examples, is available at \href{https://github.com/XukunZhu-6/Bayesian-Sequential-Logistic-Regression}{GitHub}.

\section{Simulation}\label{sec3}

\subsection{Setting}
We compare the BSLR with the ILR in a setting in which structural attrition reduces the number of subjects at risk at later transitions. Data are generated for $N = 1000$ subjects moving through a five-stage process with four sequential transitions ($S = 5$, $J = 4$). The linear predictor of each transition contains an intercept and three covariates $X_{ip} \sim N(0,1)$, $p = 1,2,3$, with transition-specific effects $\beta_{p,j}$. The covariates are drawn once and held fixed across replicates, so that the results are conditional on the realized covariate configuration.

True parameters are generated from the hierarchical structure of Section~\ref{subsec:model}: the overall effects and the transition-specific deviations are drawn independently, and the deviations are centered to satisfy the sum-to-zero constraints. The data-generating process therefore conforms to the BSLR specification, and the ILR is fitted as a less efficient competitor. Given the implied transition probabilities, outcomes are drawn as $Y_i \sim \mathrm{Categorical}(\phi_{i1},\dots,\phi_{i5})$ through the continuation-ratio factorization~\eqref{eq:cr_factorization}.

Two scenarios represent different degrees of sparsity. In the High Probability scenario most subjects advance at each transition. In the Low Probability scenario attrition concentrates at the first two transitions, so that few subjects remain at risk at transitions 3 and 4. The two scenarios were selected by drawing 100 candidate parameter sets from the process above, computing the implied cumulative probabilities $P(Y_i \ge s)$ under the fixed design matrix, and retaining the sets with the highest and the lowest probability of reaching the final stage. The true parameter values and the implied transition probabilities and at-risk counts are reported in Section~D.3 of the Supplementary Materials (Tables~D.1 and D.2).

For each scenario we generate $R = 100$ replicate datasets that differ only in the realized outcomes, and fit both models to each replicate. The ILR is assigned the same priors as the corresponding BSLR effects, so that differences in performance are attributable to the partial-pooling structure and not to the choice of prior. The prior specification and the sampler settings for both models are given in Section~D.1 of the Supplementary Materials. Performance is summarized for each parameter by the bias of the posterior mean, the posterior variance averaged across replicates, and the mean squared error (MSE) of the posterior mean; their definitions are given in Section~D.2 of the Supplementary Materials.

\subsection{Simulation results}

\begin{figure}[tbp]
    \centering
    \includegraphics[width=\textwidth]{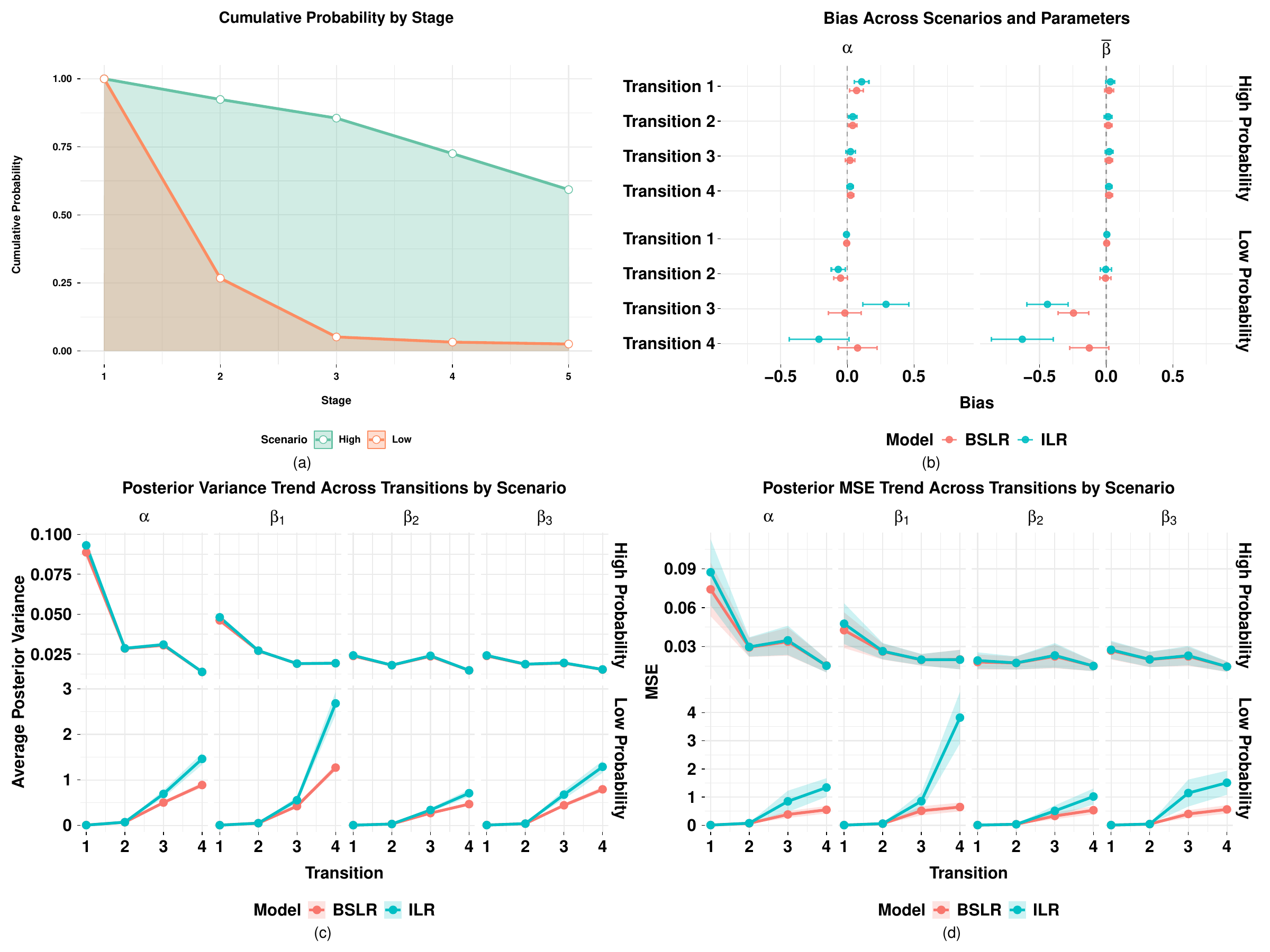}
    \caption{Simulation results under the High and Low Probability scenarios. (a) Cumulative probability of reaching each stage. (b) Bias of the posterior mean by transition for the intercept $\alpha$ and for the average of the three covariate effects $\beta$, with intervals across the $R = 100$ replicates. (c) Posterior variance averaged across replicates. (d) Mean squared error of the posterior mean, with Monte Carlo bands.}
\label{fig:sim}
\end{figure}

Figure~\ref{fig:sim} reports the results under both scenarios. Panel~(a) shows the cumulative probability of reaching each stage. Under the High Probability scenario this probability declines gradually across stages, so that a large number of subjects remain at risk at every transition. Under the Low Probability scenario it declines at the first two transitions, so that transitions 3 and 4 are estimated from a small at-risk sample.

Under the High Probability scenario the two models give bias close to zero (panel~(b)), comparable average posterior variance (panel~(c)), and overlapping MSE curves and Monte Carlo bands (panel~(d)) at all four transitions. Under the Low Probability scenario the two models agree at transitions 1 and 2 and separate from transition 3 onward. At transitions 3 and 4 the ILR bias for $\alpha$ and $\beta$ departs from zero and its intervals widen, whereas the BSLR bias remains closer to zero. The average posterior variance and the MSE of both models increase at these transitions, but the increase is smaller under the BSLR, and the ILR exceeds the BSLR for all four parameters, with the largest difference at transition 4.

The performance of the two models depends on the size of the at-risk sample. Where the at-risk sample is large, as in the High Probability scenario and at the early transitions of the Low Probability scenario, the two models give equivalent results. Where attrition has reduced the at-risk sample, the BSLR gives lower bias, lower posterior variance and lower MSE than the ILR. This difference follows from the structure of the BSLR, which estimates each transition-specific coefficient as a deviation from an effect shared by all transitions, so that the transitions with larger at-risk samples contribute information to the estimation at the transitions with fewer subjects at risk.

\subsection{Robustness to duration dependence}\label{sec:sim_robust}

\begin{figure}[tbp]
    \centering
    \includegraphics[width=\textwidth]{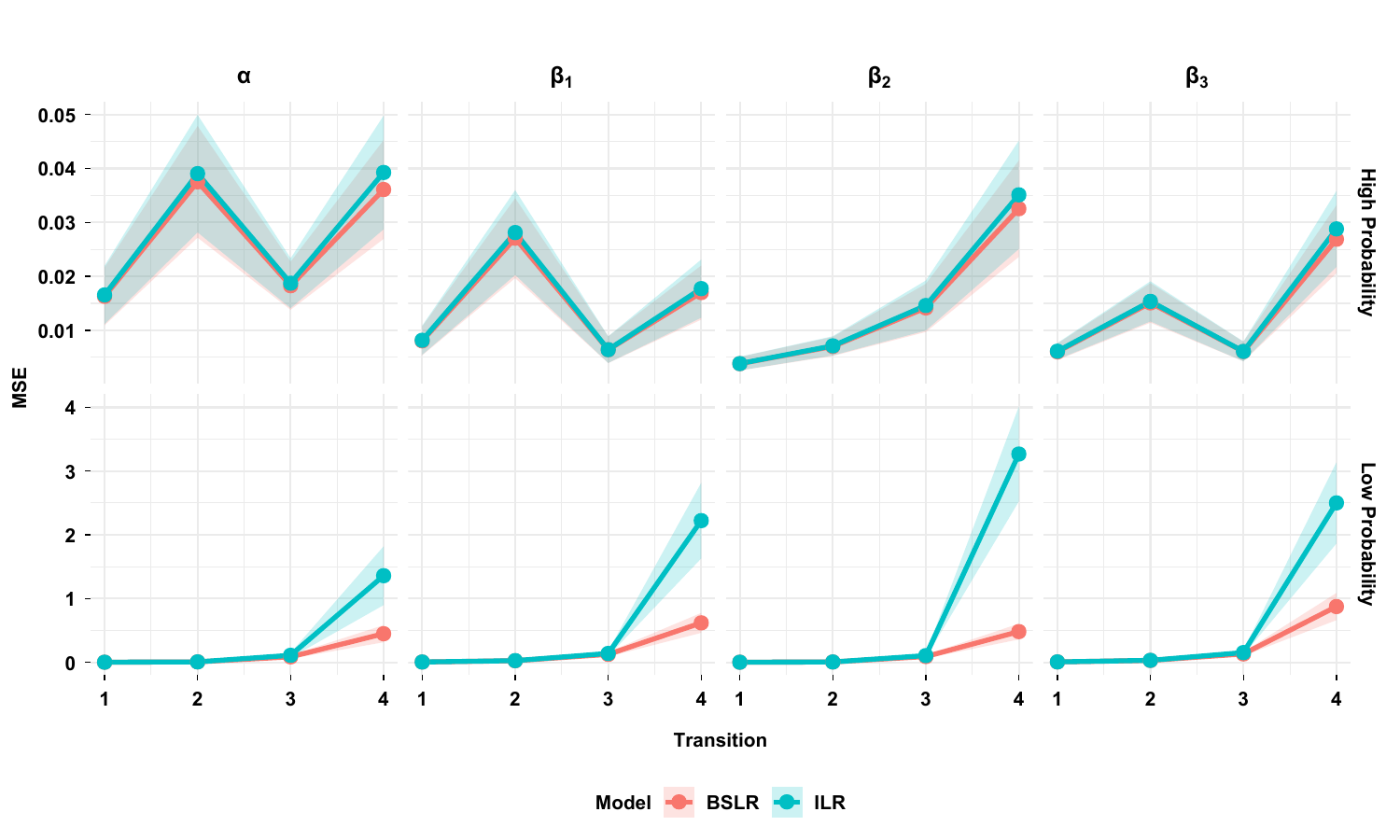}
    \caption{Mean squared error of the posterior mean under duration dependence for the intercept $\alpha$ and the covariate effects $\beta_1$, $\beta_2$ and $\beta_3$, with Monte Carlo bands. The two scenarios use separate vertical scales.}
\label{fig:sim_duration}
\end{figure}

The data in Section~3.2 are generated from the BSLR specification, under which, conditional on the covariates, the transition probability does not depend on the time a subject has spent at the current stage. In transplant evaluation this assumption may not hold: a prolonged stay within a stage may indicate a lower probability of advancing, and the subjects who remain at risk at later transitions become a selected subset of the cohort. We therefore conduct a second simulation to examine whether the results of Section~3.2 hold when the data-generating process departs from the fitted model. This study differs from the first in two respects. First, the linear predictor of the data-generating process includes a duration term. Its overall effect is negative, so that a longer stay lowers the probability of advancing, and its transition-specific components follow a monotone gradient, so that the effect is stronger at later transitions. Second, the duration term is not included in either fitted model, so that both the BSLR and the ILR are misspecified. The covariates, scenarios, priors and number of replicates are as in Section~3.1, and we report the MSE, which summarizes bias and variance jointly.The duration term and the parameter values of this study are described in Section~D.4 of the Supplementary Materials.

Figure~\ref{fig:sim_duration} reports the MSE of the posterior mean under duration dependence. Under the High Probability scenario the MSE curves and Monte Carlo bands of the two models overlap for all four parameters at all transitions. Under the Low Probability scenario the two models agree at transitions 1 to 3. At transition 4 the ILR MSE increases for all four parameters and its Monte Carlo bands widen, whereas the increase under the BSLR is smaller. Compared with Section~3.2, the two models separate one transition later.

The pattern observed in Section~3.2 is therefore retained under duration dependence. The two models give equivalent results where the at-risk sample is large, and the BSLR gives lower MSE at the transition with the fewest subjects at risk. The advantage of partial pooling is thus not limited to the case in which the data are generated from the BSLR specification.

\section{Application to CHART dataset}\label{sec4}

To demonstrate the utility of the BSLR model in a real-world clinical setting, we applied the framework to the Consortium for the Holistic Assessment of Risk in Transplant (CHART) dataset. Developed and maintained by Duke University School of Medicine, the CHART initiative was designed to leverage multidimensional health system data to improve transplant access and address data fragmentation in the organ allocation process \citep{McElroy2025}. Historically, evaluating a patient's transplant eligibility requires synthesizing an enormous volume of data scattered across semi-structured clinical notes, often failing to systematically incorporate social determinants of health (SDOH) \citep{Schappe2025}. This reliance on subjective judgments has inadvertently contributed to inequities in transplant access.

The CHART dataset overcomes these challenges by establishing a Common Data Model that seamlessly integrates longitudinal clinical electronic health record (EHR) data with multidimensional SDOH information \citep{10.1001/jamanetworkopen.2025.7704}. By reducing subjective bias and improving transparency, CHART facilitates a more equitable and holistic multidimensional risk assessment.

\begin{figure}[htbp]
    \centering
    \includegraphics[width=0.8\textwidth]{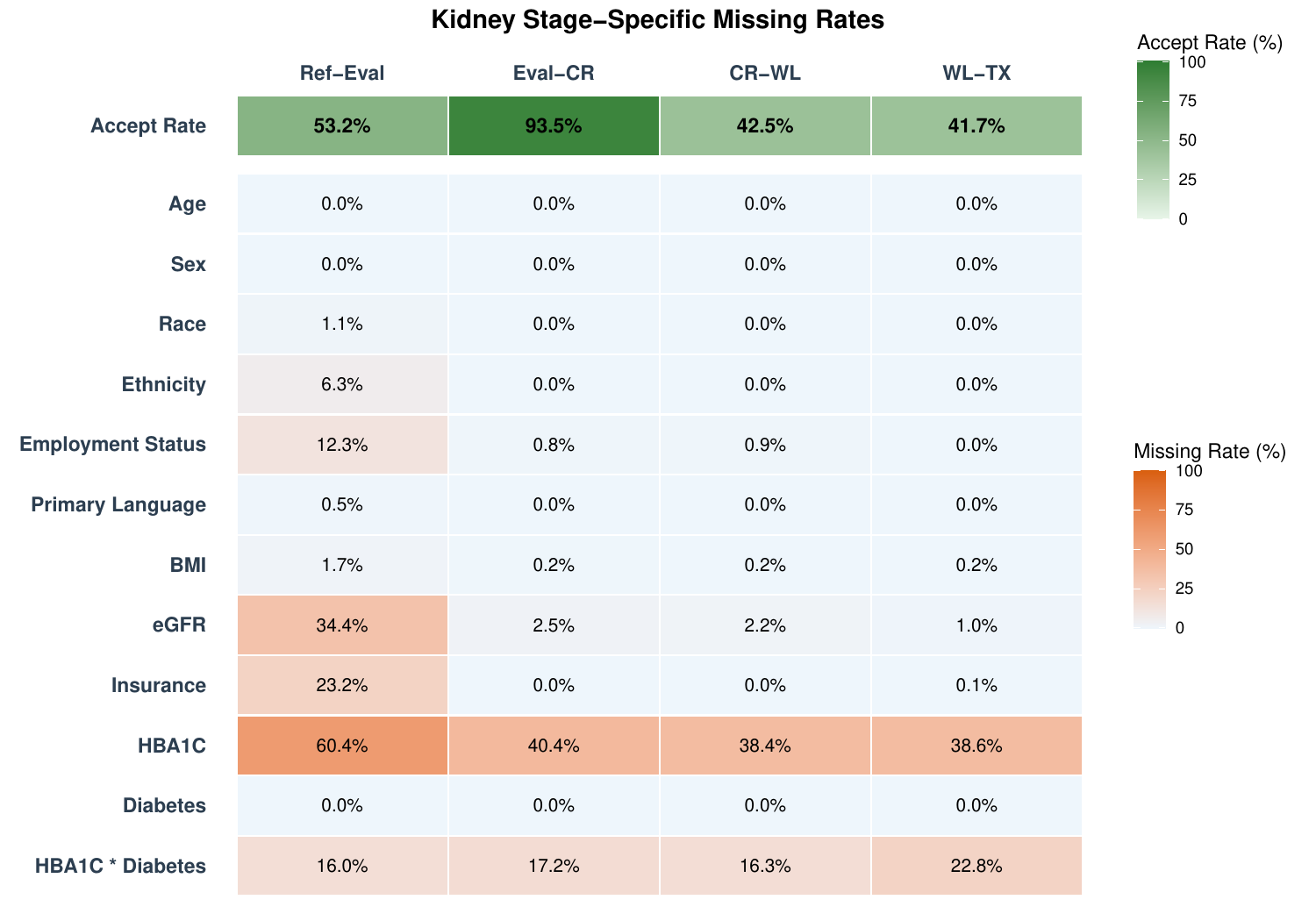}
    \caption{Stage-specific missing data rates in the kidney
    transplantation cohort.}
    \label{fig:missing_kidney}
\end{figure}

For this study, the primary analysis presented in this section focuses on the Kidney transplantation cohort ($N = 9{,}012$) extracted from the CHART database. A parallel analysis was conducted for the Liver cohort ($N = 3{,}429$), the full methodology and results of which are detailed in Section~B of the supplementary material. The multi-stage evaluation process was defined by five distinct patient states, yielding four sequential transitions: (1) Referral to Evaluation (Ref to Eval), (2) Evaluation to Committee Review (Eval to CR), (3) Committee Review to Waitlist (CR to WL), and (4) Waitlist to Transplant (WL to TX). Patient trajectories were strictly mapped to these chronological transitions.

Covariates were categorized into time-invariant and time-varying components. Time-invariant covariates include baseline demographics and SDOH factors: sex, race, ethnicity, primary language, and employment status. Time-varying covariates, which dynamically update at each transition, include patient age (years), body mass index (BMI, kg/m$^2$), and primary insurance type. To account for organ-specific clinical continuums, the kidney cohort included the estimated glomerular filtration rate (eGFR, mL/min/1.73~m$^2$), computed from serum creatinine using the race-free 2021 CKD-EPI equation \citep{inker2021egfr}, and a categorical interaction term between diabetes status and glycated hemoglobin (HbA1c, \%). Specifically, this interaction term first classifies patients by diabetes status and then stratifies those with diabetes by their HbA1c level, yielding three groups: no diabetes; controlled diabetes; and uncontrolled diabetes. Patients without a diabetes diagnosis are assigned to the first group irrespective of HbA1c.

Because eGFR is not clinically assessed during the initial administrative Referral to Evaluation step, it is structurally unavailable at that transition. We therefore set the corresponding mask entry to zero, $M_{\text{eGFR},1} = 0$, so that eGFR enters the linear predictor only at transitions 2 through 4 in accordance with the masking mechanism of Section~\ref{subsec:model}. Its overall effect is consequently an average over the three transitions at which it is measured.

The evaluation continuum exhibits distinct patterns of missing data, primarily driven by the clinical sequencing of specific tests. Figure~\ref{fig:missing_kidney} illustrates the transition missing rates for the kidney cohort. Notably, while raw HbA1c measurements demonstrate consistently high missing rates across all evaluation stages, the missing rate decreases to approximately $20\%$ after creating the interaction term with diabetes status, since patients without a diabetes diagnosis do not require an HbA1c measurement.

To handle missingness, values for covariates that persist across stages were imputed using the Last Observation Carried Forward (LOCF) method, strictly restricted by the maximum evaluation stage the patient actually reached. Remaining missing values were subsequently imputed using Multiple Imputation by Chained Equations (MICE) with predictive mean matching, ensuring that conditional dependencies were maintained. Imputation was restricted to values that were genuinely missing for patients at risk: a stage-$j$ covariate was imputed only for patients who reached stage $j$, so that no values were generated for stages a patient never entered, and the patient-stage outcome was excluded from the set of predictors. We generated $m = 5$ imputed datasets from independently seeded MICE runs, fitted the model separately to each completed dataset, and pooled the resulting posterior draws across the five analyses to form a single posterior sample. This mixture of the $m$ posteriors approximates the posterior distribution obtained by averaging over the imputation model \citep{zhou2010note}, so that imputation uncertainty is propagated into all reported credible intervals. Prior to model fitting, all continuous covariates were standardized. Baseline characteristics of the final analytic kidney cohort are summarized in Table~\ref{tab:kidney_descriptive}.

\begin{table}[tbp]
\centering
\spacingset{1}
\scriptsize
\renewcommand{\arraystretch}{1.1}
\setlength{\tabcolsep}{4pt}
\caption{Descriptive Statistics of the Kidney Patient Cohort by Transition
Stage}
\label{tab:kidney_descriptive}
\begin{tabular}{lcccc}
\toprule
\textbf{Variable} & \textbf{Ref-Eval} & \textbf{Eval-CR} & \textbf{CR-WL} & \textbf{WL-TX} \\
 & \textbf{($N = 9,012$)} & \textbf{($N = 4,799$)} & \textbf{($N = 4,486$)} & \textbf{($N = 1,905$)} \\
\midrule
Transitioned, N (\%) & 4,799 (53\%) & 4,486 (93\%) & 1,905 (42\%) & 795 (42\%) \\
Age & 55 (13) & 54 (12) & 54 (12) & 52 (12) \\
BMI & 30.2 (6.8) & 29.9 (6.0) & 29.6 (6.1) & 29.7 (5.7) \\
eGFR & 12 (10) & 11 (10) & 12 (12) & 13 (13) \\
\textbf{Sex} & & & & \\
\hspace{2mm} Female & 3,539 (39\%) & 1,860 (39\%) & 1,728 (39\%) & 747 (39\%) \\
\hspace{2mm} Male & 5,473 (61\%) & 2,939 (61\%) & 2,758 (61\%) & 1,158 (61\%) \\
\addlinespace
\textbf{Race} & & & & \\
\hspace{2mm} White & 3,047 (34\%) & 1,725 (36\%) & 1,628 (36\%) & 778 (41\%) \\
\hspace{2mm} Black & 5,217 (58\%) & 2,739 (57\%) & 2,548 (57\%) & 975 (51\%) \\
\hspace{2mm} Other & 480 (5.3\%) & 267 (5.6\%) & 248 (5.5\%) & 129 (6.8\%) \\
\hspace{2mm} Unknown & 268 (3.0\%) & 68 (1.4\%) & 62 (1.4\%) & 23 (1.2\%) \\
\addlinespace
\textbf{Ethnicity} & & & & \\
\hspace{2mm} Hispanic / Latino & 255 (2.8\%) & 146 (3.0\%) & 130 (2.9\%) & 67 (3.5\%) \\
\hspace{2mm} Not Hispanic & 7,645 (85\%) & 4,379 (91\%) & 4,097 (91\%) & 1,750 (92\%) \\
\hspace{2mm} Unknown & 1,112 (12\%) & 274 (5.7\%) & 259 (5.8\%) & 88 (4.6\%) \\
\addlinespace
\textbf{Primary Language} & & & & \\
\hspace{2mm} English & 8,785 (97\%) & 4,708 (98\%) & 4,402 (98\%) & 1,877 (99\%) \\
\hspace{2mm} Other & 227 (2.5\%) & 91 (1.9\%) & 84 (1.9\%) & 28 (1.5\%) \\
\addlinespace
\textbf{HbA1c $\times$ Diabetes} & & & & \\
\hspace{2mm} Non-Diabetic & 6,053 (67\%) & 2,409 (50\%) & 2,246 (50\%) & 967 (51\%) \\
\hspace{2mm} Controlled & 1,737 (19\%) & 1,413 (29\%) & 1,336 (30\%) & 622 (33\%) \\
\hspace{2mm} Uncontrolled & 1,222 (14\%) & 977 (20\%) & 904 (20\%) & 316 (17\%) \\
\addlinespace
\textbf{Insurance} & & & & \\
\hspace{2mm} Commercial & 1,189 (13\%) & 929 (19\%) & 810 (18\%) & 398 (21\%) \\
\hspace{2mm} Medicaid & 1,230 (14\%) & 835 (17\%) & 720 (16\%) & 306 (16\%) \\
\hspace{2mm} Medicare & 4,259 (47\%) & 2,845 (59\%) & 2,788 (62\%) & 1,126 (59\%) \\
\hspace{2mm} Other & 2,334 (26\%) & 190 (4.0\%) & 168 (3.7\%) & 75 (3.9\%) \\
\addlinespace
\textbf{Employment Status} & & & & \\
\hspace{2mm} Full Time & 1,102 (12\%) & 780 (16\%) & 722 (16\%) & 471 (25\%) \\
\hspace{2mm} Disability & 3,108 (34\%) & 2,073 (43\%) & 1,939 (43\%) & 737 (39\%) \\
\hspace{2mm} Retired & 1,853 (21\%) & 1,154 (24\%) & 1,098 (24\%) & 426 (22\%) \\
\hspace{2mm} Other & 1,096 (12\%) & 597 (12\%) & 551 (12\%) & 231 (12\%) \\
\hspace{2mm} Unknown & 1,853 (21\%) & 195 (4.1\%) & 176 (3.9\%) & 40 (2.1\%) \\
\bottomrule
\end{tabular}
\end{table}

\begin{table}[tbp]
\centering
\spacingset{1}
\scriptsize
\renewcommand{\arraystretch}{1.3}
\setlength{\tabcolsep}{3pt}
\caption{Bayesian Sequential Logistic Regression Results}
\label{tab:kidney_final_verified}
\begin{tabular}{lccccc}
\toprule
\textbf{Variables} & \textbf{Overall} & \textbf{Ref-Eval} & \textbf{Eval-CR} & \textbf{CR-WL} & \textbf{WL-TX} \\
 & \textbf{OR (95\% CI)} & \textbf{Total OR (95\% CI)} & \textbf{Total OR (95\% CI)} & \textbf{Total OR (95\% CI)} & \textbf{Total OR (95\% CI)} \\
\midrule
Age & \textbf{0.80 (0.76, 0.85)} & \textbf{0.74 (0.70, 0.80)} & 0.87 (0.76, 1.00) & \textbf{0.66 (0.61, 0.71)} & 0.95 (0.85, 1.07) \\
BMI & 0.96 (0.92, 1.00) & \textbf{0.91 (0.86, 0.97)} & 0.92 (0.82, 1.01) & 0.97 (0.91, 1.04) & 1.04 (0.94, 1.16) \\
eGFR & \textbf{1.18 (1.10, 1.28)} & - & \textbf{1.29 (1.09, 1.56)} & \textbf{0.87 (0.81, 0.93)} & \textbf{1.47 (1.32, 1.66)} \\
Female & 1.02 (0.95, 1.11) & \textbf{0.88 (0.79, 0.98)} & 0.89 (0.71, 1.10) & 1.08 (0.95, 1.24) & \textbf{1.28 (1.06, 1.56)} \\
Other Language & \textbf{0.65 (0.47, 0.93)} & \textbf{0.65 (0.45, 0.96)} & 0.87 (0.47, 2.19) & \textbf{0.51 (0.29, 0.82)} & 0.61 (0.30, 1.21) \\
\addlinespace
\multicolumn{6}{l}{\textbf{HbA1c $\times$ Diabetes (Ref: No Diabetes)}} \\
\hspace{3mm} Controlled & \textbf{1.94 (1.74, 2.16)} & \textbf{4.79 (4.16, 5.53)} & 1.15 (0.88, 1.49) & \textbf{3.13 (2.67, 3.68)} & 0.82 (0.65, 1.03) \\
\hspace{3mm} Uncontrolled & \textbf{1.66 (1.46, 1.87)} & \textbf{4.28 (3.63, 5.04)} & \textbf{1.44 (1.05, 1.99)} & \textbf{2.26 (1.84, 2.74)} & \textbf{0.54 (0.41, 0.72)} \\
\addlinespace
\multicolumn{6}{l}{\textbf{Race (Ref: White)}} \\
\hspace{3mm} Black & \textbf{0.79 (0.72, 0.87)} & \textbf{0.75 (0.66, 0.83)} & 0.86 (0.69, 1.10) & \textbf{0.69 (0.59, 0.79)} & 0.88 (0.73, 1.09) \\
\hspace{3mm} Other & 1.01 (0.85, 1.21) & 1.01 (0.81, 1.27) & 1.01 (0.71, 1.42) & 1.09 (0.86, 1.45) & 0.95 (0.67, 1.25) \\
\hspace{3mm} Unknown & 0.76 (0.54, 1.05) & \textbf{0.65 (0.44, 0.92)} & 0.78 (0.43, 1.60) & 0.75 (0.47, 1.22) & 0.85 (0.48, 1.84) \\
\addlinespace
\multicolumn{6}{l}{\textbf{Ethnicity (Ref: Non-Hispanic)}} \\
\hspace{3mm} Hispanic/Latino & 0.97 (0.76, 1.21) & 0.88 (0.63, 1.22) & 0.69 (0.36, 1.19) & 1.16 (0.80, 1.78) & 1.22 (0.77, 2.08) \\
\hspace{3mm} Unknown & \textbf{0.76 (0.63, 0.94)} & \textbf{0.68 (0.56, 0.83)} & 1.08 (0.66, 2.02) & \textbf{0.71 (0.54, 0.92)} & \textbf{0.64 (0.40, 0.94)} \\
\addlinespace
\multicolumn{6}{l}{\textbf{Insurance (Ref: Commercial)}} \\
\hspace{3mm} Medicare & \textbf{0.61 (0.54, 0.70)} & \textbf{0.52 (0.44, 0.61)} & 1.06 (0.78, 1.45) & \textbf{0.32 (0.26, 0.38)} & 0.81 (0.64, 1.04) \\
\hspace{3mm} Medicaid & \textbf{0.72 (0.63, 0.83)} & \textbf{0.70 (0.58, 0.83)} & 0.77 (0.61, 1.08) & \textbf{0.67 (0.55, 0.80)} & \textbf{0.74 (0.59, 0.96)} \\
\hspace{3mm} Other & \textbf{0.44 (0.35, 0.56)} & \textbf{0.07 (0.05, 0.08)} & 0.90 (0.50, 1.69) & \textbf{0.70 (0.49, 0.99)} & 0.92 (0.54, 1.55) \\
\addlinespace
\multicolumn{6}{l}{\textbf{Employment Status (Ref: Full Time)}} \\
\hspace{3mm} Disability & 0.95 (0.83, 1.06) & 1.01 (0.84, 1.20) & 1.00 (0.73, 1.39) & \textbf{0.64 (0.52, 0.78)} & 1.25 (0.98, 1.60) \\
\hspace{3mm} Other & \textbf{0.71 (0.61, 0.82)} & \textbf{0.74 (0.61, 0.90)} & 0.80 (0.59, 1.21) & \textbf{0.61 (0.47, 0.77)} & \textbf{0.69 (0.52, 0.92)} \\
\hspace{3mm} Retired & 1.14 (0.98, 1.35) & 1.07 (0.87, 1.30) & 1.35 (0.97, 2.14) & 1.01 (0.79, 1.26) & 1.17 (0.91, 1.58) \\
\hspace{3mm} Unknown & \textbf{0.29 (0.21, 0.39)} & \textbf{0.21 (0.17, 0.27)} & 0.61 (0.35, 1.11) & \textbf{0.28 (0.18, 0.41)} & \textbf{0.20 (0.07, 0.44)} \\
\bottomrule
\multicolumn{6}{l}{\footnotesize $^{*}$ Values presented are the odds ratios (OR). \textbf{Bold} values indicate that the 95\% credible interval excludes one.}
\end{tabular}
\end{table}

\begin{figure}[tbp]
    \centering
    \includegraphics[width=\textwidth]{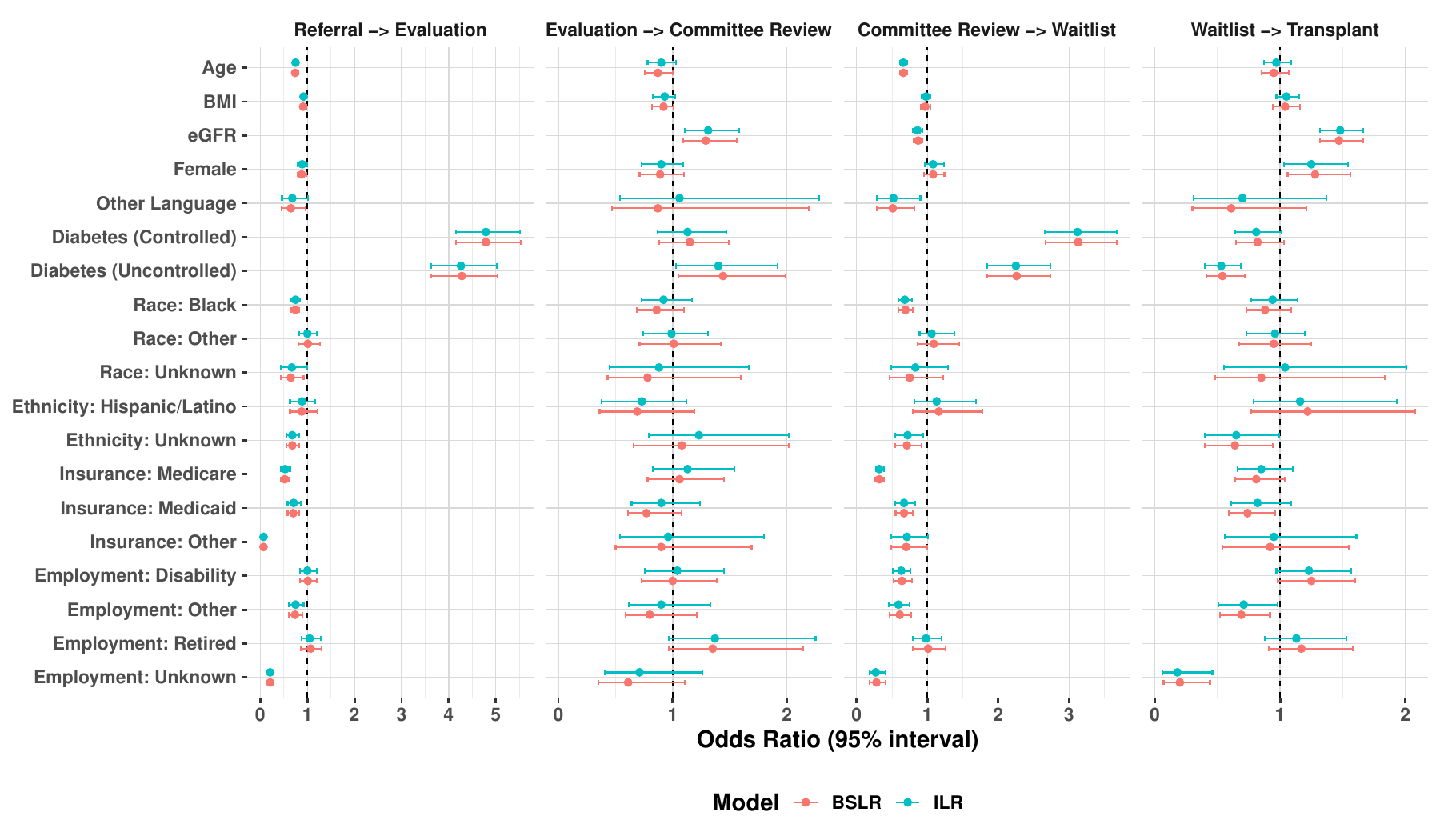}
    \caption{Stage-specific odds ratios for the kidney cohort, comparing
    the Bayesian sequential logistic regression (BSLR) and independent
    logistic regression (ILR) models across the four sequential
    transitions. The dashed vertical line at $\text{OR}=1$ indicates no
    association. Through partial pooling, BSLR yields more stable estimates
    and narrower intervals than ILR, most evidently at the later,
    lower-volume transitions.}
    \label{fig:forest_kidney}
\end{figure}

\subsection{Model specification}

Both the proposed BSLR model and the independent logistic regression were fitted to the kidney cohort (analogous models for the liver cohort are detailed in Section~B of the supplementary material). The hierarchical decomposition~\eqref{eq:decomposition}, the sum-to-zero constraints on the transition-specific deviations, and the half-Student-$t$ hyper-priors on all scale parameters are exactly as specified in Sections~\ref{subsec:model} and~\ref{subsec:computation}; no modifications were made for the application. The masking of eGFR at the first transition is the only cohort-specific feature of the specification.

Posterior inference was performed using MCMC sampling. For each model, four parallel chains were run with 100{,}000 iterations per chain, including a warm-up period of 50{,}000 iterations. Convergence was assessed using the potential scale reduction factor ($\widehat{R}$) and the effective sample size (ESS); all parameters satisfied $\widehat{R} < 1.01$ with sufficient effective sample sizes for stable posterior summaries.

The BSLR model estimates overall and transition-specific effects for the kidney cohort as summarized in Table~\ref{tab:kidney_final_verified}. Within this sequential framework, an estimated odds ratio (OR) above one indicates a higher probability of advancing to the subsequent stage, whereas a value below one denotes a reduced likelihood that acts as a barrier to progression. The results reveal that the impact of clinical and non-clinical factors shifts dynamically across the care continuum. Because all continuous covariates were standardized prior to fitting, each odds ratio for a continuous covariate corresponds to a one standard deviation increase.

\subsection{Results}

At the initial referral to evaluation transition, associations emerge across the demographic, clinical, and socioeconomic domains. Older age (OR = 0.74), higher BMI (OR = 0.91), Black race (OR = 0.75), and non-commercial insurance, such as Medicare (OR = 0.52), all correspond to lower odds of progression. Diabetes runs in the opposite direction: both controlled (OR = 4.79) and uncontrolled (OR = 4.28) diabetes are tied to much higher odds of clearing this administrative step. The picture narrows at the evaluation to committee review transition, where most demographic associations are no longer credibly different from the null and clinical measures take over, with higher eGFR (OR = 1.29) and uncontrolled diabetes (OR = 1.44) marking better odds of progression.

The committee review to waitlist transition is where disparities resurface. Older age again tracks with lower odds of advancing (OR = 0.66), and relative to their reference groups, Black candidates (OR = 0.69) and those insured through Medicaid (OR = 0.67) or Medicare (OR = 0.32) are less likely to reach the waitlist. Diabetes tells the most striking story: having been a positive predictor in earlier stages (uncontrolled OR = 2.26), it turns into a liability at the final waitlist to transplant step, where uncontrolled diabetes now carries lower odds of advancing (OR = 0.54), a shift that likely reflects the stricter surgical criteria governing final allocation. Medicaid coverage continued to be a negative predictor at this final stage (OR = 0.74), while higher eGFR (OR = 1.47) and female sex (OR = 1.28) now work in a patient's favor.

Comparing the BSLR and ILR models illustrates the advantage of borrowing strength across transitions when later stages are sparse. As shown in Figure~\ref{fig:forest_kidney}, the two models agree closely on the direction of association at every transition and give nearly identical estimates at the early, data-rich transitions. At the later, low-volume transitions, however, the ILR estimates are attenuated toward the null and have wider 95\% credible intervals, so that several associations remain inconclusive under ILR but are resolved by BSLR. For example, at the final waitlist to transplant transition, which involves the smallest subgroup in the cohort, Medicaid coverage is credibly associated with lower odds of advancing under BSLR (OR = 0.74), whereas the ILR estimate is attenuated and inconclusive (OR = 0.82). A similar pattern holds for patients with other insurance at the committee review to waitlist transition, where the association is resolved by BSLR (OR = 0.70) but not by ILR (OR = 0.71). Notably, the reverse does not occur: there is no transition at which ILR identifies an association that BSLR misses. This suggests that partial pooling makes the later, sparse transitions more precise while still keeping the real effects.

To ensure the robustness of our posterior estimates to the choice of variance priors, a sensitivity analysis was conducted. We refitted the models replacing the Half-Student-$t$ priors with Half-Cauchy priors for all variance components. The posterior estimates derived from the Half-Cauchy specification demonstrated consistency with our primary results, confirming that our clinical findings are not artifacts of prior specification. The detailed transition-specific estimates from this sensitivity analysis are provided in Section~C of the supplementary material.

\subsection{Heterogeneity Analysis}
\label{Heterogeneity Analysis}
The hierarchical structure of the BSLR framework allows us to measure how much a covariate's effect changes across the evaluation continuum. For each covariate, we estimate the posterior standard deviation $\sigma$ of its transition-specific deviation parameters, that is, the stage-level departures from the covariate's overall effect. A larger $\sigma$ means the effect fluctuates substantially between transitions. Because these deviations are on the log-odds scale, the size of $\sigma$ depends on the scale of the covariate: standardized continuous predictors are measured per unit change, whereas categorical predictors are measured relative to their reference category. The $\sigma$ estimates for the kidney cohort (Table \ref{tab:sigma_prior_comparison}) are therefore compared within, but not across, these two groups. \\

\begin{table}[htbp]
\centering
\scriptsize
\renewcommand{\arraystretch}{1.4}
\setlength{\tabcolsep}{8pt}

\caption{Posterior Estimates of Local Heterogeneity}
\label{tab:sigma_prior_comparison}

\begin{tabular}{lcc}
\toprule
\textbf{Variables} & \textbf{Half-T Prior} & \textbf{Half-Cauchy Prior} \\
\midrule

% --- Patient Characteristics ---
Age & 0.34 (0.10, 1.23) & 0.34 (0.10, 1.19) \\
BMI & 0.15 (0.02, 0.57) & 0.15 (0.02, 0.57) \\
eGFR & 0.77 (0.17, 3.01) & 0.76 (0.17, 2.95) \\
Female & 0.37 (0.09, 1.28) & 0.36 (0.09, 1.21) \\
Other Language & 0.55 (0.02, 2.07) & 0.54 (0.02, 2.02) \\

% --- HbA1c x Diabetes ---

\multicolumn{3}{l}{\textbf{HbA1c $\times$ Diabetes (Ref: No Diabetes)}} \\
\hspace{3mm} Controlled & 1.35 (0.51, 3.59) & 1.34 (0.52, 3.59) \\
\hspace{3mm} Uncontrolled & 1.40 (0.54, 3.66) & 1.40 (0.54, 3.61) \\

% --- Race ---

\multicolumn{3}{l}{\textbf{Race (Ref: White)}} \\
\hspace{3mm} Black & 0.27 (0.03, 0.99) & 0.29 (0.03, 1.08) \\
\hspace{3mm} Other & 0.23 (0.01, 1.00) & 0.22 (0.01, 0.90) \\
\hspace{3mm} Unknown & 0.41 (0.01, 1.60) & 0.41 (0.02, 1.66) \\

% --- Ethnicity ---

\multicolumn{3}{l}{\textbf{Ethnicity (Ref: Non-Hispanic)}} \\
\hspace{3mm} Hispanic/Latino & 0.57 (0.03, 2.04) & 0.55 (0.02, 1.98) \\
\hspace{3mm} Unknown & 0.49 (0.03, 1.82) & 0.48 (0.03, 1.72) \\

% --- Insurance ---

\multicolumn{3}{l}{\textbf{Insurance (Ref: Commercial)}} \\
\hspace{3mm} Medicare & 0.92 (0.31, 2.65) & 0.93 (0.32, 2.56) \\
\hspace{3mm} Medicaid & 0.20 (0.01, 0.86) & 0.20 (0.01, 0.89) \\
\hspace{3mm} Other & 1.87 (0.77, 4.53) & 1.87 (0.77, 4.70) \\

% --- Employment ---

\multicolumn{3}{l}{\textbf{Employment Status (Ref: Full Time)}} \\
\hspace{3mm} Disability & 0.56 (0.16, 1.79) & 0.54 (0.16, 1.73) \\
\hspace{3mm} Other & 0.29 (0.02, 1.14) & 0.29 (0.02, 1.13) \\
\hspace{3mm} Retired & 0.30 (0.01, 1.20) & 0.30 (0.01, 1.17) \\
\hspace{3mm} Unknown & 0.97 (0.25, 2.92) & 0.94 (0.25, 2.81) \\

\bottomrule
\multicolumn{3}{l}{\footnotesize * Values are the posterior mean of $\sigma$ (log-odds scale) and its 95\% credible interval.} 
\\
\end{tabular}
\end{table}

Among continuous predictors, eGFR shows larger cross-transition variation ($\sigma = 0.77$) than BMI ($\sigma = 0.15$), indicating that the effect of renal function is stage-dependent while the effect of body mass is more uniform. For categorical predictors, the largest variation is seen for the HbA1c $\times$ diabetes interaction (controlled $\sigma = 1.35$; uncontrolled $\sigma = 1.40$) and insurance status (Medicare $\sigma = 0.92$; other $\sigma = 1.87$), while race is more stable (Black $\sigma = 0.27$). This matches our transition-specific findings, where diabetes shifts from a facilitator in the early stages to a penalty at final allocation, and insurance type has its strongest effect at the later transitions. The heterogeneity estimates are nearly identical under the Half-Student-$t$ and Half-Cauchy priors, indicating that the inferred variation is robust to the choice between these two priors.

\section{Conclusion}\label{sec5}

In this study, we proposed a BSLR framework to model patient attrition across the multi-stage organ transplant care continuum. By reformulating the sequential progression as a series of conditional Bernoulli distributions within a joint likelihood, the model accommodates the chronological dependencies of the evaluation process. To mitigate estimation instability caused by data sparsity in later transitions, the framework utilizes a hierarchical prior structure that decomposes covariate effects into an overall effect and a transition-specific deviation. Furthermore, simulation studies demonstrate that under conditions of structural attrition, this partial pooling mechanism recovers parameters with lower bias and reduces posterior variance compared to standard ILR.

Applying the BSLR framework to the CHART dataset highlights how different covariates influencing patient attrition vary across phases. Posterior estimates indicate that while progression rates are lower at the Ref--Eval, CR--WL and WL--TX transitions than at Eval--CR. At these final allocation steps, patient progression is primarily driven by physiological severity and systemic barriers, such as Medicare or Medicaid insurance status. By separating true stage-specific covariate effects from the uncertainty caused by sample depletion, the model clarifies the clinical and socioeconomic hurdles restricting equitable organ access.

For future research, this approach can be extended in several ways. The current framework evaluates binary transitions but does not account for the time spent in each stage. Incorporating multi-state survival models or competing risk frameworks would address temporal dynamics, such as pre-waitlist mortality \citep{albert2001sequential, wiesner2001meld}. Additionally, Hidden Markov Models could be used to infer unobserved latent health states and their transitions \citep{scott2005hidden, wall2009multiple, gupta2020utilizing}. Finally, relaxing the linear additive assumption for continuous covariates by integrating non-parametric methods, such as Bayesian penalized splines, would better capture non-linear physiological trends \citep{crainiceanu2007spatially, mullah2021lasso}. In summary, this methodology offers a practical statistical tool for analyzing sequential healthcare delivery processes.
\clearpage

  \bibliography{bibliography}

@article{watson2012organ,
  title={Organ transplantation: historical perspective and current practice},
  author={Watson, CJE and Dark, JH},
  journal={British journal of anaesthesia},
  volume={108},
  number={suppl\_1},
  pages={i29--i42},
  year={2012},
  publisher={Oxford University Press}
}

@article{bezinover2019organ,
  title={Organ transplantation in the modern era},
  author={Bezinover, Dmitri and Saner, Fuat},
  journal={BMC anesthesiology},
  volume={19},
  number={1},
  pages={32},
  year={2019},
  publisher={Springer}
}

@article{Berchuck2024Predictive,
  title   = {Use of Predictive Models to Determine Transplant Eligibility},
  author  = {Berchuck, Samuel I. and Bhavsar, Nirav and Schappe, Taylor and Zaribafzadeh, Hooman and Matsouaka, Roland A. and McElroy, Linda M.},
  journal = {Current Transplantation Reports},
  year    = {2024},
  volume  = {11},
  number  = {4},
  pages   = {243--250},
  doi     = {10.1007/s40472-024-00454-4},
  pmid    = {40832452},
  pmcid   = {PMC12360406},
  note    = {Epub 2024 Oct 8}
}

@article{DeutschLink2021,
  author    = {Deutsch-Link, Sasha and Weinberg, Ethan M. and Bittermann, Therese and McDougal, Maureen and Dhariwal, Anita and Jones, Lisa S. and Weinrieb, Robert M. and Banerjee, Anurag G. and Addis, Suzanne and Serper, Marina},
  title     = {The {Stanford} Integrated Psychosocial Assessment for Transplant Is Associated With Outcomes Before and After Liver Transplantation},
  journal   = {Liver Transplantation},
  year      = {2021},
  volume    = {27},
  number    = {5},
  pages     = {652--667},
  doi       = {10.1002/lt.25975},
  pmid      = {33320417},
  pmcid     = {PMC9531321}
}

@article{10.1001/jamanetworkopen.2025.49629,
    author = {Katz-Greenberg, Goni and McElroy, Lisa M.},
    title = {Geography and Access to Kidney Transplant—Are We Measuring What Matters?},
    journal = {JAMA Network Open},
    volume = {8},
    number = {12},
    pages = {e2549629-e2549629},
    year = {2025},
    month = {12},
    issn = {2574-3805},
    doi = {10.1001/jamanetworkopen.2025.49629},
    url = {https://doi.org/10.1001/jamanetworkopen.2025.49629},
    eprint = {https://jamanetwork.com/journals/jamanetworkopen/articlepdf/2843240/katzgreenberg_2025_ic_250353_1766090498.33526.pdf},
}

@article{bertsimas2019development,
  title={Development and validation of an optimized prediction of mortality for candidates awaiting liver transplantation},
  author={Bertsimas, Dimitris and Kung, Jerry and Trichakis, Nikolaos and Wang, Yuchen and Hirose, Ryutaro and Vagefi, Parsia A},
  journal={American Journal of Transplantation},
  volume={19},
  number={4},
  pages={1109--1118},
  year={2019},
  publisher={Wiley Online Library}
}

@article{ascher2024chalf,
  title={CHALF Score: a novel tool to rapidly risk stratify children in need of liver transplant evaluation during acute liver failure},
  author={Ascher-Bartlett, Johanna M and Bangerth, Sarah and Jordan, Shannon and Weaver, Carly and Barhouma, Sarah and Etesami, Kambiz and Kohli, Rohit and Emamaullee, Juliet},
  journal={Transplantation},
  volume={108},
  number={4},
  pages={930--939},
  year={2024},
  publisher={LWW}
}

@article{zhang2024development,
  title={Development and validation of a risk score predicting death without transplant in adult heart transplant candidates},
  author={Zhang, Kevin C and Narang, Nikhil and Jasseron, Carine and Dorent, Richard and Lazenby, Kevin A and Belkin, Mark N and Grinstein, Jonathan and Mayampurath, Anoop and Churpek, Matthew M and Khush, Kiran K and others},
  journal={JAMA},
  volume={331},
  number={6},
  pages={500--509},
  year={2024}
}

@article{Lee2010,
  author    = {Lee, Changmin and Ayers, Stephanie L. and Kronenfeld, Jennie J. and others},
  title     = {The importance of examining movements within the {US} health care system: sequential logit modeling},
  journal   = {BMC Health Services Research},
  volume    = {10},
  number    = {1},
  pages     = {269},
  year      = {2010},
  doi       = {10.1186/1472-6963-10-269},
  url       = {https://doi.org/10.1186/1472-6963-10-269}
}

@article{rigon2019bayesian,
  title={Bayesian semiparametric modelling of contraceptive behaviour in India via sequential logistic regressions},
  author={Rigon, Tommaso and Durante, Daniele and Torelli, Nicola},
  journal={Journal of the Royal Statistical Society Series A: Statistics in Society},
  volume={182},
  number={1},
  pages={225--247},
  year={2019},
  publisher={Oxford University Press}
}

@article{wideman2009prospective,
  title={A prospective sequential analysis of the fear-avoidance model of pain},
  author={Wideman, Timothy H and Adams, Heather and Sullivan, Michael JL},
  journal={PAIN{\textregistered}},
  volume={145},
  number={1-2},
  pages={45--51},
  year={2009},
  publisher={Elsevier}
}

@article{dissanayake2002analysis,
  title={Analysis of severity of young driver crashes: Sequential binary logistic regression modeling},
  author={Dissanayake, Sunanda and Lu, John},
  journal={Transportation Research Record},
  volume={1784},
  number={1},
  pages={108--114},
  year={2002},
  publisher={SAGE Publications Sage CA: Los Angeles, CA}
}

@article{deppen2014predicting,
  title={Predicting lung cancer prior to surgical resection in patients with lung nodules},
  author={Deppen, Stephen A and Blume, Jeffrey D and Aldrich, Melinda C and Fletcher, Sarah A and Massion, Pierre P and Walker, Ronald C and Chen, Heidi C and Speroff, Theodore and Degesys, Catherine A and Pinkerman, Rhonda and others},
  journal={Journal of Thoracic Oncology},
  volume={9},
  number={10},
  pages={1477--1484},
  year={2014},
  publisher={Elsevier}
}

@book{powers2008statistical,
  title={Statistical methods for categorical data analysis},
  author={Powers, Daniel and Xie, Yu},
  year={2008},
  publisher={Emerald Group Publishing}
}

@book{agresti2010analysis,
  title={Analysis of ordinal categorical data},
  author={Agresti, Alan},
  year={2010},
  publisher={John Wiley \& Sons}
}

@article{carpenter2017stan,
  title   = {Stan: A Probabilistic Programming Language},
  author  = {Carpenter, Bob and Gelman, Andrew and Hoffman, Matthew D. and Lee, Daniel and Goodrich, Ben and Betancourt, Michael and Brubaker, Marcus A. and Guo, Jiqiang and Li, Peter and Riddell, Allen},
  journal = {Journal of Statistical Software},
  volume  = {76},
  number  = {1},
  pages   = {1--32},
  year    = {2017}
}

@article{gelman2015stan,
  title   = {Stan: A Probabilistic Programming Language for Bayesian Inference and Optimization},
  author  = {Gelman, Andrew and Lee, Daniel and Guo, Jiqiang},
  journal = {Journal of Educational and Behavioral Statistics},
  volume  = {40},
  number  = {5},
  pages   = {530--543},
  year    = {2015}
}

@article{gelman2006prior,
  title   = {Prior Distributions for Variance Parameters in Hierarchical Models},
  author  = {Gelman, Andrew},
  journal = {Bayesian Analysis},
  volume  = {1},
  number  = {3},
  pages   = {515--534},
  year    = {2006}
}

@article{hoffman2014nuts,
  title   = {The No-U-Turn Sampler: Adaptively Setting Path Lengths in Hamiltonian Monte Carlo},
  author  = {Hoffman, Matthew D. and Gelman, Andrew},
  journal = {Journal of Machine Learning Research},
  volume  = {15},
  pages   = {1593--1623},
  year    = {2014}
}

@article{polson2012half,
  title   = {On the Half-Cauchy Prior for a Global Scale Parameter},
  author  = {Polson, Nicholas G. and Scott, James G.},
  journal = {Bayesian Analysis},
  volume  = {7},
  number  = {4},
  pages   = {887--902},
  year    = {2012}
}

@article{vehtari2021rank,
  title   = {Rank-Normalization, Folding, and Localization: An Improved $\widehat{R}$ for Assessing Convergence of MCMC},
  author  = {Vehtari, Aki and Gelman, Andrew and Simpson, Daniel and Carpenter, Bob and B\"urkner, Paul-Christian},
  journal = {Bayesian Analysis},
  volume  = {16},
  number  = {2},
  pages   = {667--718},
  year    = {2021}
}

@book{hoff2009first,
  title={A first course in Bayesian statistical methods},
  author={Hoff, Peter D},
  volume={580},
  year={2009},
  publisher={Springer}
}

@article{seltzer1996bayesian,
  title={Bayesian analysis in applications of hierarchical models: Issues and methods},
  author={Seltzer, Michael H and Wong, Wing Hung and Bryk, Anthony S},
  journal={Journal of Educational and Behavioral Statistics},
  volume={21},
  number={2},
  pages={131--167},
  year={1996},
  publisher={Sage Publications Sage CA: Los Angeles, CA}
}

@article{McElroy2025,
  author    = {McElroy, Lisa M. and Reed, Rhiannon D. and Gordon, Elisa J. and Strauss, Alexandra T. and Harding, Jessica and Adams, Abigail and Caicedo, Juan Carlos and Ross Driscoll, Kimberly and Taber, David J. and Ng, Yue Harn and Bhavsar, Nrupen A. and Patzer, Rachel E. and Kirk, Allan D.},
  title     = {Consortium for the Holistic Assessment of Risk in Transplant: Harmonizing Data for Research, Transparency, and Equity},
  journal   = {Annals of Surgery},
  year      = {2025},
  volume    = {281},
  number    = {3},
  pages     = {373--375},
  doi       = {10.1097/SLA.0000000000006410},
  pmid      = {38899463},
  pmcid     = {PMC11659506}
}

@article{Schappe2025,
  author    = {Schappe, Tyler and McElroy, Lisa M. and Ogundolie, Mobolaji and Matsouaka, Roland and Rogers, Ursula and Bhavsar, Nrupen A.},
  title     = {A data pipeline for secure extraction and sharing of social determinants of health},
  journal   = {PLoS ONE},
  year      = {2025},
  volume    = {20},
  number    = {1},
  pages     = {e0317215},
  doi       = {10.1371/journal.pone.0317215},
  pmid      = {39888883},
  pmcid     = {PMC11785280}
}

@article{10.1001/jamanetworkopen.2025.7704,
    author = {McElroy, Lisa M. and Rogers, Ursula and Nichols, Liz and Bhavsar, Nrupen A. and Schappe, Tyler and Harding, Jessica and Strauss, Alexandra T. and Gordon, Elisa J. and Ross-Driscoll, Katie and Deirhoi-Reed, Rhiannon and Chan, Norine W. and Henson, Jackie B. and Ng, Yue Harn and Taber, Dave J. and Milam, Adam J. and Schold, Jesse D. and Caicedo, Juan Carlos and Adams, Andrew B. and Patzer, Rachel E. and Berchuck, Samuel I. and Matsouaka, Roland and Gagnon, Jennifer and Kirk, Allan D.},
    title = {Development of Common Data Elements for Organ Transplantation},
    journal = {JAMA Network Open},
    volume = {8},
    number = {4},
    pages = {e257704-e257704},
    year = {2025},
    month = {04},
    issn = {2574-3805},
    doi = {10.1001/jamanetworkopen.2025.7704},
    url = {https://doi.org/10.1001/jamanetworkopen.2025.7704},
    eprint = {https://jamanetwork.com/journals/jamanetworkopen/articlepdf/2833164/mcelroy_2025_ld_250043_1744818970.84429.pdf},
}

@article{scott2005hidden,
  title={Hidden Markov models for longitudinal comparisons},
  author={Scott, Steven L and James, Gareth M and Sugar, Catherine A},
  journal={Journal of the American Statistical Association},
  volume={100},
  number={470},
  pages={359--369},
  year={2005},
  publisher={Taylor \& Francis}
}

@article{wall2009multiple,
  title={Multiple indicator hidden Markov model with an application to medical utilization data},
  author={Wall, Melanie M and Li, Ran},
  journal={Statistics in medicine},
  volume={28},
  number={2},
  pages={293--310},
  year={2009},
  publisher={Wiley Online Library}
}

@article{gupta2020utilizing,
  title={Utilizing time series data embedded in electronic health records to develop continuous mortality risk prediction models using hidden Markov models: a sepsis case study},
  author={Gupta, Akash and Liu, Tieming and Crick, Christopher},
  journal={Statistical methods in medical research},
  volume={29},
  number={11},
  pages={3409--3423},
  year={2020},
  publisher={SAGE Publications Sage UK: London, England}
}

@article{wiesner2001meld,
  title={MELD and PELD: application of survival models to liver allocation},
  author={Wiesner, Russell H and McDiarmid, Sue V and Kamath, Patrick S and Edwards, Eric B and Malinchoc, Michael and Kremers, Walter K and Krom, Ruud AF and Kim, W Ray},
  journal={Liver transplantation},
  volume={7},
  number={7},
  pages={567--580},
  year={2001},
  publisher={Wiley Online Library}
}

@article{albert2001sequential,
  title={Sequential ordinal modeling with applications to survival data},
  author={Albert, James H and Chib, Siddhartha},
  journal={Biometrics},
  volume={57},
  number={3},
  pages={829--836},
  year={2001},
  publisher={Oxford University Press}
}

@article{crainiceanu2007spatially,
  title={Spatially adaptive Bayesian penalized splines with heteroscedastic errors},
  author={Crainiceanu, Ciprian M and Ruppert, David and Carroll, Raymond J and Joshi, Adarsh and Goodner, Billy},
  journal={Journal of Computational and Graphical Statistics},
  volume={16},
  number={2},
  pages={265--288},
  year={2007},
  publisher={Taylor \& Francis}
}

@article{mullah2021lasso,
  title={LASSO type penalized spline regression for binary data},
  author={Mullah, Muhammad Abu Shadeque and Hanley, James A and Benedetti, Andrea},
  journal={BMC medical research methodology},
  volume={21},
  number={1},
  pages={83},
  year={2021},
  publisher={Springer}
}

@article{zhou2010note,
  title   = {A Note on Bayesian Inference After Multiple Imputation},
  author  = {Zhou, Xiang and Reiter, Jerome P.},
  journal = {The American Statistician},
  volume  = {64},
  number  = {2},
  pages   = {159--163},
  year    = {2010}
}

@article{inker2021egfr,
  title   = {New Creatinine- and Cystatin C-Based Equations to Estimate GFR without Race},
  author  = {Inker, Lesley A. and Eneanya, Nwamaka D. and Coresh, Josef and Tighiouart, Hocine and Wang, Dan and Sang, Yingying and Crews, Deidra C. and Doria, Alessandro and Estrella, Michelle M. and Froissart, Marc and others},
  journal = {New England Journal of Medicine},
  volume  = {385},
  number  = {19},
  pages   = {1737--1749},
  year    = {2021}
}

@article{tutz1991sequential,
  author  = {Tutz, Gerhard},
  title   = {Sequential Models in Categorical Regression},
  journal = {Computational Statistics \& Data Analysis},
  volume  = {11},
  number  = {3},
  pages   = {275--295},
  year    = {1991},
  issn    = {0167-9473},
  doi     = {10.1016/0167-9473(91)90086-H}
}

\clearpage
\section*{Appendix A: Detailed Derivations}
\label{app:derivations}
\setcounter{equation}{0}
\renewcommand{\theequation}{A.\arabic{equation}}

This appendix supplies the derivations deferred from Section~2.2: the equivalence
of the transition probabilities (Section~A.1), the continuation-ratio
factorization of the cell probabilities (Section~A.2), the verification of the
compact likelihood representation (Section~A.3), and the resulting equality of
the two sample likelihoods (Section~A.4).

\subsection*{Preliminaries}

There are $S$ sequential stages and $J = S - 1$ transitions. The categorical
endpoint $Y_i \in \{1, \ldots, S\}$ denotes the final stage reached by patient
$i$, with cell probabilities $\phi_{is} = P(Y_i = s)$, and transition $j$ denotes
the attempted move from stage $j$ to stage $j+1$. The transition indicators are
\begin{equation}
W_{ij} = \mathbb{I}(Y_i \ge j+1), \qquad j = 0, 1, \ldots, J,
\label{eq:app_deterministic}
\end{equation}
where the definition is stated from $j = 0$ so that
$W_{i0} = \mathbb{I}(Y_i \ge 1) \equiv 1$ holds automatically rather than by
convention; only $j = 1, \ldots, J$ index modelled transitions. For every
$j = 1, \ldots, J$,
\begin{equation}
\{W_{i,j-1} = 1\} \;=\; \{Y_i \ge j\} \;=\; \{\text{patient } i \text{ reached stage } j\},
\label{eq:app_atrisk}
\end{equation}
so that $W_{i,j-1}$ indicates whether patient $i$ was at risk for transition $j$;
at $j = 1$ both events are certain, since every patient is at risk for the first
transition.\\

Throughout, all probabilities are conditional on $\theta$, on $Z_i$ and on a covariate path $X_{i,1:J}=(X_{i,1},\dots,X_{i,J})$, which is treated as fixed; this conditioning is suppressed for readability. Under this conditioning $Y_i$ has a single well-defined distribution on $\{1,\dots,S\}$, so that $\phi_{i1},\dots,\phi_{iS}$ are the cells of one distribution. Since, under the specification of Section~2.3, $\omega_{ij}$ depends on the path only through $X_{i,j}$, the likelihood contribution of patient $i$ depends on the path only through $X_{i,1:\min(y_i,J)}$, by the truncation noted in Section~A.3; components at stages the patient never reached do not enter the likelihood and need not be observed or defined. We assume $P(Y_i\ge j)>0$ for $j=1,\dots,J$, so that every conditional probability below is well defined; under the logit specification of Section~2.3 we moreover have $0<\omega_{ij}<1$.

\subsection*{A.1 Equivalence of the transition probabilities}

The Bernoulli transition probability is
$\pi_{ij} = P(W_{ij} = 1 \mid W_{i,j-1} = 1)$, $j = 1, \ldots, J$. By
\eqref{eq:app_deterministic}, $\{W_{ij} = 1\} = \{Y_i \ge j+1\}$ and
$\{W_{i,j-1} = 1\} = \{Y_i \ge j\}$, and since
$\{Y_i \ge j+1\} \subseteq \{Y_i \ge j\}$,
\begin{equation}
P(W_{ij} = 1, W_{i,j-1} = 1) = P(W_{ij} = 1),
\qquad\text{hence}\qquad
\pi_{ij} = \frac{P(W_{ij} = 1)}{P(W_{i,j-1} = 1)}.
\label{eq:app_transition_ratio}
\end{equation}
The two probabilities in \eqref{eq:app_transition_ratio} are obtained by
conditioning on the categorical endpoint. From \eqref{eq:app_deterministic},
\[
P(W_{ij} = 1 \mid Y_i = s) =
\begin{cases}
1, & s \ge j+1,\\
0, & s \le j,
\end{cases}
\]
so that the law of total probability truncates the sums,
\begin{align}
P(W_{ij} = 1) &= \sum_{s=1}^{S} P(W_{ij} = 1 \mid Y_i = s)\, \phi_{is}
= \sum_{s=j+1}^{S} \phi_{is} = P(Y_i \ge j+1),
\label{eq:app_numerator}\\
P(W_{i,j-1} = 1) &= \sum_{s=1}^{S} P(W_{i,j-1} = 1 \mid Y_i = s)\, \phi_{is}
= \sum_{s=j}^{S} \phi_{is} = P(Y_i \ge j),
\label{eq:app_denominator}
\end{align}
where at $j=1$, \eqref{eq:app_denominator} reduces to $P(Y_i \ge 1) = 1$, consistent
with $W_{i0} \equiv 1$. Substituting \eqref{eq:app_numerator} and
\eqref{eq:app_denominator} into \eqref{eq:app_transition_ratio} gives
\begin{equation}
\pi_{ij} = \frac{P(Y_i \ge j+1)}{P(Y_i \ge j)} = P(Y_i \ge j+1 \mid Y_i \ge j) = \omega_{ij},
\qquad j = 1, \ldots, J.
\label{eq:app_transition_equivalence}
\end{equation}
The Bernoulli transition probability and the continuation ratio therefore
coincide at every transition.

\subsection*{A.2 Continuation-ratio factorization of the cell probabilities}

Rearranging the definition of $\omega_{ij}$ gives the recursion
$P(Y_i \ge j+1) = P(Y_i \ge j)\, \omega_{ij}$. Starting from $P(Y_i \ge 1) = 1$
and iterating,
\begin{equation}
P(Y_i \ge s) = \prod_{l=1}^{s-1} \omega_{il}, \qquad s = 1, \ldots, S,
\label{eq:app_cumulative}
\end{equation}
where the product is taken over transitions $l$ and equals one when $s = 1$. For
$s < S$, the event $\{Y_i = s\}$ is the difference of the nested events
$\{Y_i \ge s\}$ and $\{Y_i \ge s+1\}$, so that \eqref{eq:app_cumulative} gives
\begin{equation}
\phi_{is} = P(Y_i \ge s) - P(Y_i \ge s+1)
= \Bigg(\prod_{l=1}^{s-1} \omega_{il}\Bigg)(1 - \omega_{is}),
\qquad s < S.
\label{eq:app_cell_nonterminal}
\end{equation}
By the labelling convention of Section~2.1, $\omega_{is}$ is the probability of transition $s$, which exists because $s<S$ implies $s\le J$. For the terminal stage, completing the process requires completing all $J$
transitions, so
\begin{equation}
\phi_{iS} = P(Y_i = S) = P(Y_i \ge S) = \prod_{l=1}^{J} \omega_{il}.
\label{eq:app_cell_terminal}
\end{equation}
Summing \eqref{eq:app_cell_nonterminal} over $s = 1, \ldots, S-1$ and adding
\eqref{eq:app_cell_terminal} telescopes to
$\sum_{s=1}^{S} \phi_{is} = P(Y_i \ge 1) = 1$, so the continuation-ratio
factorization defines a proper distribution over the $S$ stages.

\subsection*{A.3 Verification of the compact likelihood representation}

We show that the categorical contribution admits the unified form
\begin{equation}
\Pi_i \;\equiv\; \prod_{j=1}^{J}
\left[ \omega_{ij}^{W_{ij}} (1 - \omega_{ij})^{1 - W_{ij}} \right]^{W_{i,j-1}}
\;=\; \phi_{i, y_i}.
\label{eq:app_categorical_compact}
\end{equation}

\noindent\textit{Case 1: $y_i = s$ with $s < S$.}
By \eqref{eq:app_deterministic}, $W_{ij} = \mathbb{I}(j \le s-1)$ and
$W_{i,j-1} = \mathbb{I}(j \le s)$. The outer exponent therefore deactivates every
factor with $j \ge s+1$. Of the remaining factors, those with
$j = 1, \ldots, s-1$ have $W_{ij} = 1$ and reduce to $\omega_{ij}$, while the
factor at $j = s$ has $W_{is} = 0$ and reduces to $1 - \omega_{is}$. Hence
\begin{equation}
\Pi_i = \Bigg(\prod_{j=1}^{s-1} \omega_{ij}\Bigg)(1 - \omega_{is}) = \phi_{is},
\label{eq:app_case1}
\end{equation}
by \eqref{eq:app_cell_nonterminal}. For $s = 1$ the empty product equals one,
giving $\phi_{i1} = 1 - \omega_{i1}$.\\

\noindent\textit{Case 2: $y_i = S$.}
Here $W_{ij} = 1$ for all $j = 1, \ldots, J$ and $W_{i,j-1} = 1$ for all
$j = 1, \ldots, J$, so every exponent equals one and every factor reduces to
$\omega_{ij}$. Hence
\begin{equation}
\Pi_i = \prod_{j=1}^{J} \omega_{ij} = \phi_{iS},
\label{eq:app_case2}
\end{equation}
by \eqref{eq:app_cell_terminal}.\\
Because the outer exponent vanishes for every $j > y_i$, the product in~\eqref{eq:app_categorical_compact} truncates at $j = \min(y_i, J)$; every base is strictly positive because $0<\omega_{ij}<1$, so no convention for $0^0$ is required, and the likelihood involves $\omega_{ij}$ only for $j \le y_i$. The two cases are mutually exclusive and exhaust the support $\{1, \ldots, S\}$ of $Y_i$, which establishes \eqref{eq:app_categorical_compact}. Moreover, the map $y \mapsto \big(\mathbb{I}(y\ge 2),\dots,\mathbb{I}(y\ge S)\big)$ is a bijection from $\{1, \ldots, S\}$ onto the set of non-increasing binary vectors in $\{0,1\}^{J}$, that is, those of the form $(1, \ldots, 1, 0, \ldots, 0)$: the categorical endpoint and the realized sequence of transition outcomes carry the same information, so the two formulations are likelihoods for the same observed data.

\subsection*{A.4 Equality of the two sample likelihoods}

\textbf{Lemma A.1 (chain-rule factorization).}
For every patient $i$ and every non-increasing sequence
$1\ge w_1\ge\cdots\ge w_J\ge 0$ in $\{0,1\}^J$, with $w_0:=1$, matching $W_{i0}\equiv 1$,
\begin{equation}
P(W_{i1}=w_1,\dots,W_{iJ}=w_J)
  =\prod_{j=1}^{J}
  \Big[\pi_{ij}^{\,w_j}(1-\pi_{ij})^{1-w_j}\Big]^{w_{j-1}}.
\label{eq:app_chain}
\end{equation}

\noindent\emph{Proof.} By the chain rule,
\[
P(W_{i1}=w_1,\dots,W_{iJ}=w_J)
  =\prod_{j=1}^{J}
   P\big(W_{ij}=w_j \,\big|\, W_{i1}=w_1,\dots,W_{i,j-1}=w_{j-1}\big),
\]
where each conditioning event is of the form $\{Y_i\ge j\}$ or $\{Y_i=s\}$
for some $s<j$, and has positive probability because $0<\omega_{ij}<1$.
We evaluate the $j$-th factor in two cases.
If $w_{j-1}=1$, monotonicity gives $w_1=\dots=w_{j-1}=1$, and
$\{W_{i1}=1,\dots,W_{i,j-1}=1\}=\{Y_i\ge j\}=\{W_{i,j-1}=1\}$
by~\eqref{eq:app_atrisk}; the factor is therefore
$P(W_{ij}=w_j\mid W_{i,j-1}=1)=\pi_{ij}^{\,w_j}(1-\pi_{ij})^{1-w_j}$.
If $w_{j-1}=0$, monotonicity gives $w_j=0$, and the conditioning event is
contained in $\{W_{i,j-1}=0\}=\{Y_i\le j-1\}\subseteq\{Y_i\le j\}
=\{W_{ij}=0\}$, so the factor equals one. In both cases the factor equals
$\big[\pi_{ij}^{\,w_j}(1-\pi_{ij})^{1-w_j}\big]^{w_{j-1}}$. 

\medskip
Under the sequential Bernoulli formulation, the contribution of patient $i$
is the probability of the observed transition sequence. Since $(W_{i1},\dots,W_{iJ})$ is non-increasing and $W_{i0}\equiv1$,
Lemma~A.1 gives
\begin{equation}
L_i^{(B)}(\theta) = \prod_{j=1}^{J}
\left[ \pi_{ij}^{W_{ij}} (1 - \pi_{ij})^{1 - W_{ij}} \right]^{W_{i,j-1}},
\label{eq:app_bernoulli}
\end{equation}
in which, by~\eqref{eq:app_atrisk}, the outer exponent $W_{i,j-1}$ retains
the $j$-th Bernoulli factor only for patients who reached stage $j$ and were therefore at risk for transition $j$. Substituting the transition-probability equivalence
\eqref{eq:app_transition_equivalence} into \eqref{eq:app_bernoulli} and applying
\eqref{eq:app_categorical_compact},
\begin{equation}
L_i^{(B)}(\theta) = \prod_{j=1}^{J}
\left[ \omega_{ij}^{W_{ij}} (1 - \omega_{ij})^{1 - W_{ij}} \right]^{W_{i,j-1}}
= \phi_{i, y_i} = L_i^{(C)}(\theta).
\label{eq:app_individual_equivalence}
\end{equation}
Assuming that patients are independent conditional on their covariates and
$\theta$,
\begin{equation}
L^{(B)}(\theta) = \prod_{i=1}^{N} L_i^{(B)}(\theta)
= \prod_{i=1}^{N} L_i^{(C)}(\theta) = L^{(C)}(\theta).
\label{eq:app_sample_equivalence}
\end{equation}
The categorical continuation-ratio formulation and the sequential Bernoulli
formulation therefore induce exactly the same sample likelihood. The identity~\eqref{eq:app_individual_equivalence} can also be read directly from the bijection noted at the end of Section~A.3: since $Y_i$ and $(W_{i1},\dots,W_{iJ})$ determine each other,
$P(Y_i=y_i)=P(W_{i1}=w_1,\dots,W_{iJ}=w_J)$ with $w_j=\mathbb{I}(y_i\ge j+1)$,
and Lemma~A.1 evaluates the right-hand side. The individual-level identity follows from~\eqref{eq:app_deterministic} and the law of total probability alone, holds for any specification of the probabilities $\omega_{ij}\in(0,1)$, and is unaffected by the masking mechanism; the sample-level identity~\eqref{eq:app_sample_equivalence} additionally uses conditional independence across patients, which is common to both formulations. Because the identity carries no modelling content, it is the regression specification of Section~2.3, in which $\omega_{ij}$ depends on the patient's history only through $X_{i,j}$ and $Z_i$, and not on the time already spent in stage $j$ -- that carries the modelling content; robustness to this assumption is assessed in Section~3.3. If the covariate process is itself influenced by earlier progression, \eqref{eq:app_sample_equivalence} is to be read as a likelihood conditional on the observed covariate history, valid under the assumption that the covariate process carries no information about $\theta$.

\clearpage

\section*{Appendix B: Liver Transplantation Cohort Analysis}
\setcounter{table}{0}
\renewcommand{\thetable}{B.\arabic{table}}
\setcounter{figure}{0}
\renewcommand{\thefigure}{B.\arabic{figure}}

This appendix provides the descriptive statistics and transition-specific modeling results for the liver transplantation cohort (\(N = 3{,}429\)) extracted from the CHART dataset. The statistical framework, including the Bayesian Sequential Logistic Regression (BSLR) model specification, prior distributions, and the handling of missing data by LOCF followed by MICE with $m = 5$ imputations and posterior pooling, strictly mirrors the methodology detailed in the main text.

% ==========================================
% Table B.1: Liver Descriptive Statistics
% ==========================================
\begin{table}[!htbp]
\spacingset{1}
\centering
\footnotesize
\renewcommand{\arraystretch}{1.15}
\setlength{\tabcolsep}{3pt}
\caption{Descriptive Statistics of the Liver Patient Cohort by Transition Stage}
\label{tab:liver_descriptive}
\begin{tabular*}{\textwidth}{@{\extracolsep{\fill}}lcccc@{}}
\toprule
\textbf{Variable} & \textbf{Ref-Eval} & \textbf{Eval-CR} & \textbf{CR-WL} & \textbf{WL-TX} \\
 & ($N = 3{,}429$) & ($N = 1{,}593$) & ($N = 1{,}517$) & ($N = 757$) \\
\midrule
Transitioned, $N$ (\%) & 1,593 (46\%) & 1,517 (95\%) & 757 (50\%) & 586 (77\%) \\
Age  & 56 (12)    & 56 (12)    & 56 (12)    & 56 (11)    \\
BMI  & 29.8 (6.8) & 29.5 (6.1) & 29.4 (6.1) & 29.3 (5.7) \\
MELD & --         & --         & --         & 25 (8)     \\
\addlinespace
\multicolumn{5}{@{}l}{\textbf{Sex}} \\
\hspace{2mm} Female & 1,453 (42\%) & 632 (40\%) & 597 (39\%) & 280 (37\%) \\
\hspace{2mm} Male   & 1,976 (58\%) & 961 (60\%) & 920 (61\%) & 477 (63\%) \\
\addlinespace
\multicolumn{5}{@{}l}{\textbf{Race}} \\
\hspace{2mm} White   & 2,369 (69\%) & 1,290 (81\%) & 1,229 (81\%) & 628 (83\%) \\
\hspace{2mm} Black   & 337 (9.8\%)  & 169 (11\%)   & 159 (10\%)   & 68 (9.0\%) \\
\hspace{2mm} Other   & 155 (4.5\%)  & 72 (4.5\%)   & 68 (4.5\%)   & 41 (5.4\%) \\
\hspace{2mm} Unknown & 568 (17\%)   & 62 (3.9\%)   & 61 (4.0\%)   & 20 (2.6\%) \\
\addlinespace
\multicolumn{5}{@{}l}{\textbf{Ethnicity}} \\
\hspace{2mm} Hispanic / Latino & 91 (2.7\%)   & 41 (2.6\%)   & 39 (2.6\%)   & 25 (3.3\%) \\
\hspace{2mm} Not Hispanic      & 2,675 (78\%) & 1,420 (89\%) & 1,352 (89\%) & 687 (91\%) \\
\hspace{2mm} Unknown           & 663 (19\%)   & 132 (8.3\%)  & 126 (8.3\%)  & 45 (5.9\%) \\
\addlinespace
\multicolumn{5}{@{}l}{\textbf{Primary Language}} \\
\hspace{2mm} English & 3,023 (88\%) & 1,569 (98\%) & 1,496 (99\%) & 743 (98\%) \\
\hspace{2mm} Other   & 406 (12\%)   & 24 (1.5\%)   & 21 (1.4\%)   & 14 (1.8\%) \\
\addlinespace
\multicolumn{5}{@{}l}{\textbf{Insurance}} \\
\hspace{2mm} Commercial & 836 (24\%)   & 492 (31\%) & 464 (31\%) & 267 (35\%) \\
\hspace{2mm} Medicaid   & 690 (20\%)   & 404 (25\%) & 355 (23\%) & 190 (25\%) \\
\hspace{2mm} Medicare   & 1,192 (35\%) & 626 (39\%) & 632 (42\%) & 272 (36\%) \\
\hspace{2mm} Other      & 711 (21\%)   & 71 (4.5\%) & 66 (4.4\%) & 28 (3.7\%) \\
\addlinespace
\multicolumn{5}{@{}l}{\textbf{Employment Status}} \\
\hspace{2mm} Full Time  & 581 (17\%) & 317 (20\%) & 304 (20\%) & 174 (23\%) \\
\hspace{2mm} Disability & 756 (22\%) & 504 (32\%) & 482 (32\%) & 216 (29\%) \\
\hspace{2mm} Retired    & 886 (26\%) & 514 (32\%) & 491 (32\%) & 260 (34\%) \\
\hspace{2mm} Other      & 440 (13\%) & 218 (14\%) & 202 (13\%) & 99 (13\%)  \\
\hspace{2mm} Unknown    & 766 (22\%) & 40 (2.5\%) & 38 (2.5\%) & 8 (1.1\%)  \\
\bottomrule
\end{tabular*}
\end{table}

\subsection*{Missing Data and Baseline Characteristics}
\begin{figure}[!htbp]
\centering
\includegraphics[width=0.8\textwidth]{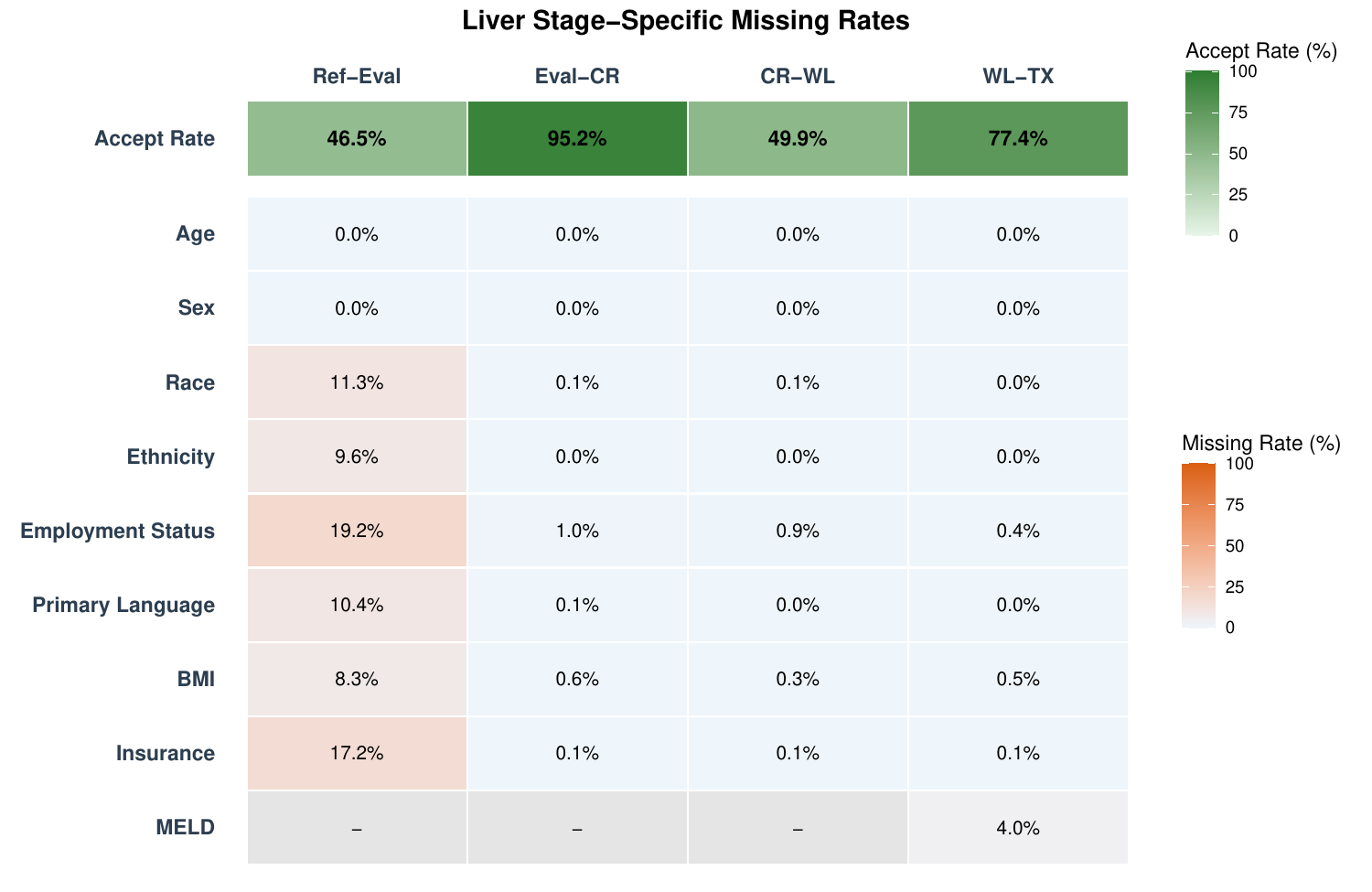}
\caption{Stage-specific missing data rates and acceptance rates for covariates in the liver transplantation cohort, highlighting the absence of MELD scores in the early administrative stages and elevated missingness of socio-demographic covariates at referral.}
\label{fig:missing_liver}
\end{figure}

Missing data in the liver cohort follows the clinical workflow closely. As shown in Figure \ref{fig:missing_liver}, the MELD score is absent throughout the early administrative stages (Referral, Evaluation, and Committee Review), since it is used for surgical urgency and organ allocation rather than initial screening. It only enters the data at the Waitlist to Transplant stage, where just 4.0\% of values are missing. The opposite pattern holds for social demographic and administrative variables, which are least complete at referral: Employment Status (19.2\%), Insurance (17.2\%), Race (11.3\%), Primary Language (10.4\%), and Ethnicity (9.6\%) all show notable gaps at this point, but fall below 1\% once patients move into active evaluation. Table \ref{tab:liver_descriptive} reports the baseline characteristics across all transitions.

\subsection*{Results}

The evaluation process for liver transplantation is uniquely driven by clinical urgency. As shown in the sequential modeling results, disease severity, quantified by the MELD score, functions as the strongest predictor of progression to transplantation (\(\text{OR} = 2.07\)), confirming that physiological urgency fundamentally dictates the final allocation step.

However, the sequential model reveals that demographic and systemic disparities are not confined to the later allocation stages but are present from the earliest point of the evaluation continuum. Female candidates face a significantly reduced likelihood of transition at the initial Referral to Evaluation stage (\(\text{OR} = 0.74\)), and this disadvantage persists through waitlist approval (\(\text{OR} = 0.80\)) and transplantation (\(\text{OR} = 0.71\)). Similarly, patients relying on Medicare exhibit consistently lower odds of progression across multiple transitions, including Referral to Evaluation (\(\text{OR} = 0.58\)), Evaluation to Committee (\(\text{OR} = 0.58\)), and Committee to Waitlist (\(\text{OR} = 0.47\)). These findings indicate that while clinical severity governs the final transplantation decision, non-clinical disparities accumulate throughout the entire process rather than emerging only at the later stages.

Comparing the BSLR and ILR models reinforces the methodological advantage of the sequential framework when modeling processes with high patient attrition. The ILR model frequently produces unstable estimates with extremely wide credible intervals for categorical variables during transitions with low event rates, a variance inflation that the BSLR framework effectively resolves via partial pooling.

% ==========================================
% Table B.2: Liver BSLR Results
% ==========================================
\begin{table}[!htbp]
\spacingset{1}
\centering
\scriptsize
\renewcommand{\arraystretch}{1.2}
\setlength{\tabcolsep}{1pt}
\caption{Bayesian Sequential Logistic Regression Results for the Liver Cohort}
\label{tab:liver_sequential}
\begin{tabular*}{\textwidth}{@{\extracolsep{\fill}}lccccc@{}}
\toprule
\textbf{Variables} & \textbf{Overall} & \textbf{Ref-Eval} & \textbf{Eval-CR} & \textbf{CR-WL} & \textbf{WL-TX} \\
 & OR (95\% CI) & OR (95\% CI) & OR (95\% CI) & OR (95\% CI) & OR (95\% CI) \\
\midrule
Age & 0.95 (0.87, 1.04) & \textbf{0.89 (0.81, 0.99)} & 0.97 (0.81, 1.17) & 0.97 (0.87, 1.10) & 0.99 (0.84, 1.20) \\
BMI & 0.97 (0.89, 1.05) & \textbf{0.89 (0.82, 0.96)} & 1.02 (0.87, 1.30) & 0.98 (0.89, 1.08) & 0.98 (0.85, 1.15) \\
Female & \textbf{0.75 (0.65, 0.87)} & \textbf{0.74 (0.64, 0.85)} & 0.77 (0.59, 1.10) & \textbf{0.80 (0.67, 0.97)} & \textbf{0.71 (0.52, 0.91)} \\
MELD Score & \textbf{2.07 (1.72, 2.52)} & -- & -- & -- & \textbf{2.07 (1.72, 2.52)} \\
Other Language & 0.91 (0.49, 1.60) & \textbf{0.50 (0.29, 0.85)} & 0.50 (0.14, 1.67) & 1.39 (0.53, 3.80) & 1.94 (0.47, 12.25) \\
\addlinespace
\multicolumn{6}{@{}l}{\textbf{Race (Ref: White)}} \\
\hspace{2mm} Black & \textbf{0.75 (0.61, 0.95)} & \textbf{0.76 (0.62, 0.95)} & 0.79 (0.54, 1.26) & \textbf{0.75 (0.57, 0.99)} & 0.72 (0.43, 1.08) \\
\hspace{2mm} Other & 1.07 (0.79, 1.52) & 0.98 (0.68, 1.38) & 1.06 (0.53, 2.22) & 1.08 (0.71, 1.75) & 1.15 (0.67, 2.47) \\
\hspace{2mm} Unknown & 0.76 (0.48, 1.26) & \textbf{0.55 (0.38, 0.80)} & 1.70 (0.50, 12.04) & 0.57 (0.32, 1.01) & 0.64 (0.25, 1.55) \\
\addlinespace
\multicolumn{6}{@{}l}{\textbf{Ethnicity (Ref: Non-Hispanic)}} \\
\hspace{2mm} Hispanic/Latino & 1.22 (0.81, 2.01) & 0.95 (0.57, 1.47) & 1.17 (0.46, 3.39) & 1.41 (0.81, 2.86) & 1.43 (0.63, 4.10) \\
\hspace{2mm} Unknown & 0.82 (0.62, 1.07) & \textbf{0.76 (0.60, 0.99)} & 0.95 (0.54, 1.99) & \textbf{0.65 (0.44, 0.92)} & 0.94 (0.57, 1.88) \\
\addlinespace
\multicolumn{6}{@{}l}{\textbf{Insurance (Ref: Commercial)}} \\
\hspace{2mm} Medicare & \textbf{0.58 (0.47, 0.71)} & \textbf{0.58 (0.47, 0.70)} & \textbf{0.58 (0.37, 0.89)} & \textbf{0.47 (0.36, 0.61)} & 0.72 (0.48, 1.15) \\
\hspace{2mm} Medicaid & 0.92 (0.76, 1.08) & 0.96 (0.79, 1.16) & 0.93 (0.64, 1.33) & 0.93 (0.74, 1.16) & 0.88 (0.59, 1.15) \\
\hspace{2mm} Other & \textbf{0.38 (0.27, 0.56)} & \textbf{0.26 (0.19, 0.35)} & 0.44 (0.23, 1.14) & \textbf{0.48 (0.29, 0.82)} & \textbf{0.37 (0.19, 0.78)} \\
\addlinespace
\multicolumn{6}{@{}l}{\textbf{Employment Status (Ref: Full Time)}} \\
\hspace{2mm} Disability & \textbf{1.28 (1.01, 1.61)} & \textbf{2.11 (1.67, 2.69)} & 1.28 (0.73, 2.25) & 0.91 (0.68, 1.21) & 1.09 (0.67, 1.76) \\
\hspace{2mm} Other & 1.01 (0.80, 1.27) & 0.99 (0.78, 1.26) & 0.80 (0.42, 1.30) & 0.90 (0.64, 1.21) & 1.46 (0.86, 2.92) \\
\hspace{2mm} Retired & \textbf{1.48 (1.13, 1.91)} & \textbf{1.83 (1.41, 2.39)} & 1.47 (0.81, 2.50) & \textbf{1.52 (1.09, 2.07)} & 1.18 (0.66, 1.95) \\
\hspace{2mm} Unknown & \textbf{0.30 (0.17, 0.56)} & \textbf{0.14 (0.10, 0.21)} & 0.85 (0.23, 4.52) & \textbf{0.26 (0.12, 0.57)} & 0.24 (0.06, 1.01) \\
\bottomrule
\end{tabular*}

\smallskip
\parbox{\textwidth}{\scriptsize OR: odds ratio; CI: credible interval. Bold values indicate that the 95\% credible interval excludes 1.}
\end{table}

% ==========================================
% Table B.3: Liver ILR Results
% ==========================================
\begin{table}[!htbp]
\spacingset{1}
\centering
\scriptsize
\renewcommand{\arraystretch}{1.2}
\setlength{\tabcolsep}{3pt}
\caption{Independent Logistic Regression Results for the Liver Cohort}
\label{tab:liver_simple}
\begin{tabular*}{\textwidth}{@{\extracolsep{\fill}}lcccc@{}}
\toprule
\textbf{Variables} & \textbf{Ref-Eval} & \textbf{Eval-CR} & \textbf{CR-WL} & \textbf{WL-TX} \\
 & OR (95\% CI) & OR (95\% CI) & OR (95\% CI) & OR (95\% CI) \\
\midrule
Age & 0.91 (0.81, 1.00) & 1.00 (0.84, 1.19) & 1.00 (0.89, 1.11) & 1.01 (0.87, 1.19) \\
BMI & \textbf{0.89 (0.82, 0.97)} & 1.03 (0.89, 1.23) & 0.99 (0.90, 1.08) & 0.99 (0.86, 1.14) \\
Female & \textbf{0.74 (0.63, 0.87)} & 0.87 (0.58, 1.28) & 0.85 (0.70, 1.05) & 0.72 (0.50, 1.01) \\
MELD Score & -- & -- & -- & \textbf{2.06 (1.71, 2.50)} \\
Other Language & \textbf{0.52 (0.30, 0.90)} & 0.55 (0.15, 1.69) & 1.41 (0.61, 3.71) & 1.87 (0.58, 11.05) \\
\addlinespace
\multicolumn{5}{@{}l}{\textbf{Race (Ref: White)}} \\
\hspace{2mm} Black & 0.81 (0.63, 1.02) & 0.92 (0.55, 1.49) & 0.80 (0.57, 1.09) & 0.82 (0.48, 1.24) \\
\hspace{2mm} Other & 0.97 (0.71, 1.28) & 1.03 (0.61, 1.89) & 1.07 (0.77, 1.64) & 1.09 (0.72, 2.08) \\
\hspace{2mm} Unknown & \textbf{0.55 (0.38, 0.81)} & 1.95 (0.68, 10.38) & 0.58 (0.33, 1.01) & 0.64 (0.24, 1.56) \\
\addlinespace
\multicolumn{5}{@{}l}{\textbf{Ethnicity (Ref: Non-Hispanic)}} \\
\hspace{2mm} Hispanic/Latino & 0.91 (0.60, 1.31) & 1.05 (0.50, 2.83) & 1.28 (0.80, 2.56) & 1.25 (0.67, 3.88) \\
\hspace{2mm} Unknown & 0.79 (0.60, 1.03) & 1.01 (0.57, 1.91) & \textbf{0.67 (0.44, 0.99)} & 1.04 (0.60, 1.88) \\
\addlinespace
\multicolumn{5}{@{}l}{\textbf{Insurance (Ref: Commercial)}} \\
\hspace{2mm} Medicare & \textbf{0.58 (0.47, 0.72)} & 0.65 (0.38, 1.08) & \textbf{0.48 (0.36, 0.62)} & 0.91 (0.60, 1.38) \\
\hspace{2mm} Medicaid & 0.99 (0.84, 1.15) & 1.00 (0.74, 1.32) & 0.97 (0.79, 1.15) & 0.96 (0.69, 1.18) \\
\hspace{2mm} Other & \textbf{0.26 (0.19, 0.35)} & 0.65 (0.27, 1.71) & \textbf{0.57 (0.34, 0.94)} & 0.51 (0.22, 1.17) \\
\addlinespace
\multicolumn{5}{@{}l}{\textbf{Employment Status (Ref: Full Time)}} \\
\hspace{2mm} Disability & \textbf{2.10 (1.64, 2.67)} & 1.23 (0.72, 2.14) & 0.88 (0.66, 1.17) & 0.92 (0.59, 1.45) \\
\hspace{2mm} Other & 0.99 (0.79, 1.24) & 0.83 (0.44, 1.21) & 0.89 (0.64, 1.15) & 1.26 (0.87, 2.43) \\
\hspace{2mm} Retired & \textbf{1.81 (1.36, 2.38)} & 1.31 (0.74, 2.43) & \textbf{1.43 (1.03, 2.01)} & 0.88 (0.54, 1.41) \\
\hspace{2mm} Unknown & \textbf{0.14 (0.10, 0.21)} & 1.20 (0.34, 5.77) & \textbf{0.28 (0.12, 0.60)} & 0.25 (0.05, 1.05) \\
\bottomrule
\end{tabular*}

\smallskip
\parbox{\textwidth}{\scriptsize OR: odds ratio; CI: credible interval. Bold values indicate that the 95\% credible interval excludes 1.}
\end{table}

% ==========================================
% Table B.4: Liver Univariable Results
% ==========================================
\begin{table}[!htbp]
\spacingset{1}
\centering
\scriptsize
\renewcommand{\arraystretch}{1.2}
\setlength{\tabcolsep}{3pt}
\caption{Univariable Logistic Regression Results for the Liver Cohort}
\label{tab:liver_univariate_new}
\begin{tabular*}{\textwidth}{@{\extracolsep{\fill}}lcccc@{}}
\toprule
\textbf{Variables} & \textbf{Ref-Eval} & \textbf{Eval-CR} & \textbf{CR-WL} & \textbf{WL-TX} \\
 & OR (95\% CI) & OR (95\% CI) & OR (95\% CI) & OR (95\% CI) \\
\midrule
Age & 0.95 (0.89, 1.01) & 0.97 (0.85, 1.13) & 0.96 (0.89, 1.04) & 0.94 (0.82, 1.04) \\
BMI & \textbf{0.92 (0.86, 0.98)} & 1.00 (0.86, 1.21) & 0.98 (0.89, 1.08) & 0.99 (0.87, 1.16) \\
Female & \textbf{0.81 (0.72, 0.92)} & 0.81 (0.62, 1.07) & \textbf{0.81 (0.70, 0.96)} & \textbf{0.80 (0.63, 0.99)} \\
MELD Score & -- & -- & -- & \textbf{2.29 (1.85, 2.86)} \\
Other Language & \textbf{0.06 (0.04, 0.09)} & 0.46 (0.14, 1.86) & 2.09 (0.85, 5.34) & 4.52 (0.82, 40.44) \\
\addlinespace
\multicolumn{5}{@{}l}{\textbf{Race (Ref: White)}} \\
\hspace{2mm} Black & 0.83 (0.68, 1.02) & 0.87 (0.56, 1.57) & \textbf{0.76 (0.57, 0.99)} & 0.86 (0.57, 1.56) \\
\hspace{2mm} Other & 0.77 (0.56, 1.06) & 0.95 (0.46, 2.06) & 1.28 (0.83, 2.15) & 1.36 (0.75, 2.98) \\
\hspace{2mm} Unknown & \textbf{0.10 (0.08, 0.14)} & 3.08 (0.62, 29.99) & \textbf{0.49 (0.28, 0.82)} & 0.56 (0.24, 1.46) \\
\addlinespace
\multicolumn{5}{@{}l}{\textbf{Ethnicity (Ref: Non-Hispanic)}} \\
\hspace{2mm} Hispanic/Latino & 0.79 (0.51, 1.18) & 1.10 (0.42, 3.39) & 1.48 (0.85, 2.79) & 1.60 (0.72, 4.80) \\
\hspace{2mm} Unknown & \textbf{0.22 (0.18, 0.27)} & 1.02 (0.48, 2.54) & \textbf{0.55 (0.37, 0.83)} & 0.89 (0.46, 1.82) \\
\addlinespace
\multicolumn{5}{@{}l}{\textbf{Insurance (Ref: Commercial)}} \\
\hspace{2mm} Medicare & \textbf{0.71 (0.60, 0.85)} & 0.68 (0.47, 1.01) & \textbf{0.56 (0.44, 0.70)} & \textbf{0.68 (0.50, 0.94)} \\
\hspace{2mm} Medicaid & 0.97 (0.81, 1.16) & 0.93 (0.65, 1.28) & 0.94 (0.75, 1.16) & 0.93 (0.69, 1.22) \\
\hspace{2mm} Other & \textbf{0.08 (0.06, 0.11)} & 0.55 (0.23, 1.51) & 0.62 (0.37, 1.03) & \textbf{0.43 (0.20, 0.96)} \\
\addlinespace
\multicolumn{5}{@{}l}{\textbf{Employment Status (Ref: Full Time)}} \\
\hspace{2mm} Disability & \textbf{1.59 (1.32, 1.95)} & 1.09 (0.67, 1.67) & \textbf{0.62 (0.51, 0.84)} & 1.00 (0.61, 1.32) \\
\hspace{2mm} Other & \textbf{0.80 (0.65, 0.99)} & 0.62 (0.41, 1.11) & 0.82 (0.59, 1.05) & 1.24 (0.74, 2.36) \\
\hspace{2mm} Retired & 1.13 (0.92, 1.33) & 1.02 (0.71, 1.52) & 0.89 (0.72, 1.14) & 0.82 (0.55, 1.11) \\
\hspace{2mm} Unknown & \textbf{0.05 (0.03, 0.06)} & 0.73 (0.25, 3.56) & \textbf{0.22 (0.10, 0.42)} & \textbf{0.20 (0.04, 0.60)} \\
\bottomrule
\end{tabular*}

\smallskip
\parbox{\textwidth}{\scriptsize OR: odds ratio; CI: credible interval. Bold values indicate that the 95\% credible interval excludes 1.}
\end{table}

\clearpage
\section*{Appendix C: Sensitivity Analysis Results}
\setcounter{table}{0}
\renewcommand{\thetable}{C.\arabic{table}}

% ==========================================
% Table C.1: Kidney BSLR - Half-Cauchy
% ==========================================
\begin{table}[!htbp]
\spacingset{1}
\centering
\scriptsize
\renewcommand{\arraystretch}{1.2}
\setlength{\tabcolsep}{1pt}
\caption{Bayesian Sequential Logistic Regression Results for the Kidney Cohort: Half-Cauchy Prior}
\label{tab:kidney_sequential_cauchy}
\begin{tabular*}{\textwidth}{@{\extracolsep{\fill}}lccccc@{}}
\toprule
\textbf{Variables} & \textbf{Overall} & \textbf{Ref-Eval} & \textbf{Eval-CR} & \textbf{CR-WL} & \textbf{WL-TX} \\
 & OR (95\% CI) & OR (95\% CI) & OR (95\% CI) & OR (95\% CI) & OR (95\% CI) \\
\midrule
Age & \textbf{0.80 (0.76, 0.84)} & \textbf{0.74 (0.70, 0.79)} & 0.87 (0.76, 1.01) & \textbf{0.66 (0.60, 0.71)} & 0.95 (0.85, 1.07) \\
BMI & 0.96 (0.92, 1.00) & \textbf{0.91 (0.86, 0.97)} & 0.92 (0.82, 1.01) & 0.97 (0.92, 1.04) & 1.04 (0.94, 1.16) \\
eGFR & \textbf{1.18 (1.10, 1.28)} & -- & \textbf{1.29 (1.10, 1.56)} & \textbf{0.86 (0.81, 0.93)} & \textbf{1.47 (1.31, 1.65)} \\
Female & 1.02 (0.95, 1.11) & \textbf{0.88 (0.79, 0.99)} & 0.89 (0.72, 1.10) & 1.09 (0.96, 1.23) & \textbf{1.28 (1.05, 1.55)} \\
Other Language & \textbf{0.65 (0.46, 0.92)} & \textbf{0.65 (0.45, 0.95)} & 0.87 (0.47, 2.17) & \textbf{0.51 (0.29, 0.81)} & 0.61 (0.29, 1.17) \\
\addlinespace
\multicolumn{6}{@{}l}{\textbf{HbA1c $\times$ Diabetes (Ref: No Diabetes)}} \\
\hspace{2mm} Controlled & \textbf{1.92 (1.73, 2.12)} & \textbf{4.81 (4.18, 5.56)} & 1.13 (0.87, 1.48) & \textbf{3.11 (2.66, 3.64)} & 0.80 (0.63, 1.00) \\
\hspace{2mm} Uncontrolled & \textbf{1.65 (1.45, 1.87)} & \textbf{4.27 (3.64, 5.02)} & \textbf{1.43 (1.05, 2.00)} & \textbf{2.25 (1.87, 2.70)} & \textbf{0.54 (0.40, 0.71)} \\
\addlinespace
\multicolumn{6}{@{}l}{\textbf{Race (Ref: White)}} \\
\hspace{2mm} Black & \textbf{0.79 (0.72, 0.87)} & \textbf{0.75 (0.67, 0.83)} & 0.86 (0.69, 1.10) & \textbf{0.69 (0.59, 0.79)} & 0.89 (0.73, 1.09) \\
\hspace{2mm} Other & 1.01 (0.85, 1.21) & 1.01 (0.80, 1.27) & 1.01 (0.71, 1.42) & 1.08 (0.86, 1.44) & 0.96 (0.67, 1.31) \\
\hspace{2mm} Unknown & 0.75 (0.54, 1.06) & \textbf{0.65 (0.45, 0.92)} & 0.78 (0.44, 1.61) & 0.75 (0.47, 1.21) & 0.85 (0.48, 1.88) \\
\addlinespace
\multicolumn{6}{@{}l}{\textbf{Ethnicity (Ref: Non-Hispanic)}} \\
\hspace{2mm} Hispanic/Latino & 0.96 (0.76, 1.21) & 0.88 (0.63, 1.21) & 0.70 (0.36, 1.17) & 1.16 (0.80, 1.77) & 1.21 (0.78, 2.07) \\
\hspace{2mm} Unknown & \textbf{0.76 (0.63, 0.93)} & \textbf{0.68 (0.56, 0.83)} & 1.07 (0.67, 2.00) & \textbf{0.71 (0.54, 0.92)} & \textbf{0.64 (0.41, 0.95)} \\
\addlinespace
\multicolumn{6}{@{}l}{\textbf{Insurance (Ref: Commercial)}} \\
\hspace{2mm} Medicare & \textbf{0.61 (0.54, 0.70)} & \textbf{0.52 (0.44, 0.61)} & 1.05 (0.78, 1.44) & \textbf{0.32 (0.26, 0.38)} & 0.81 (0.64, 1.04) \\
\hspace{2mm} Medicaid & \textbf{0.72 (0.63, 0.83)} & \textbf{0.70 (0.58, 0.83)} & 0.77 (0.60, 1.08) & \textbf{0.67 (0.55, 0.80)} & \textbf{0.74 (0.60, 0.97)} \\
\hspace{2mm} Other & \textbf{0.45 (0.35, 0.56)} & \textbf{0.07 (0.05, 0.08)} & 0.90 (0.50, 1.70) & \textbf{0.71 (0.49, 0.99)} & 0.92 (0.54, 1.55) \\
\addlinespace
\multicolumn{6}{@{}l}{\textbf{Employment Status (Ref: Full Time)}} \\
\hspace{2mm} Disability & 0.95 (0.83, 1.07) & 1.01 (0.85, 1.20) & 1.00 (0.73, 1.38) & \textbf{0.64 (0.52, 0.78)} & 1.25 (0.98, 1.60) \\
\hspace{2mm} Other & \textbf{0.71 (0.61, 0.82)} & \textbf{0.73 (0.61, 0.89)} & 0.80 (0.59, 1.20) & \textbf{0.61 (0.47, 0.76)} & \textbf{0.70 (0.53, 0.92)} \\
\hspace{2mm} Retired & 1.15 (0.98, 1.40) & 1.07 (0.88, 1.30) & 1.36 (0.98, 2.14) & 1.02 (0.80, 1.26) & 1.18 (0.91, 1.69) \\
\hspace{2mm} Unknown & \textbf{0.29 (0.21, 0.40)} & \textbf{0.21 (0.17, 0.27)} & 0.61 (0.35, 1.10) & \textbf{0.28 (0.19, 0.41)} & \textbf{0.20 (0.07, 0.51)} \\
\bottomrule
\end{tabular*}

\smallskip
\parbox{\textwidth}{\scriptsize OR: odds ratio; CI: credible interval. Bold values indicate that the 95\% credible interval excludes 1.}
\end{table}

% ==========================================
% Table C.2: Kidney ILR - Half-Cauchy
% ==========================================
\begin{table}[!htbp]
\spacingset{1}
\centering
\scriptsize
\renewcommand{\arraystretch}{1.2}
\setlength{\tabcolsep}{3pt}
\caption{Independent Logistic Regression Results for the Kidney Cohort: Half-Cauchy Prior}
\label{tab:kidney_simple_cauchy}
\begin{tabular*}{\textwidth}{@{\extracolsep{\fill}}lcccc@{}}
\toprule
\textbf{Variables} & \textbf{Ref-Eval} & \textbf{Eval-CR} & \textbf{CR-WL} & \textbf{WL-TX} \\
 & OR (95\% CI) & OR (95\% CI) & OR (95\% CI) & OR (95\% CI) \\
\midrule
Age & \textbf{0.75 (0.70, 0.80)} & 0.90 (0.77, 1.03) & \textbf{0.66 (0.61, 0.72)} & 0.97 (0.87, 1.09) \\
BMI & \textbf{0.92 (0.87, 0.97)} & 0.93 (0.83, 1.02) & 0.98 (0.92, 1.04) & 1.05 (0.96, 1.15) \\
eGFR & -- & \textbf{1.31 (1.11, 1.58)} & \textbf{0.86 (0.80, 0.93)} & \textbf{1.48 (1.32, 1.67)} \\
Female & \textbf{0.89 (0.80, 0.99)} & 0.90 (0.73, 1.10) & 1.08 (0.96, 1.23) & \textbf{1.25 (1.03, 1.53)} \\
Other Language & 0.68 (0.46, 1.00) & 1.06 (0.54, 2.25) & \textbf{0.52 (0.29, 0.89)} & 0.68 (0.30, 1.35) \\
\addlinespace
\multicolumn{5}{@{}l}{\textbf{HbA1c $\times$ Diabetes (Ref: No Diabetes)}} \\
\hspace{2mm} Controlled & \textbf{4.81 (4.17, 5.54)} & 1.12 (0.87, 1.46) & \textbf{3.11 (2.65, 3.65)} & \textbf{0.78 (0.62, 0.98)} \\
\hspace{2mm} Uncontrolled & \textbf{4.25 (3.56, 5.00)} & \textbf{1.40 (1.03, 1.93)} & \textbf{2.25 (1.87, 2.70)} & \textbf{0.53 (0.40, 0.69)} \\
\addlinespace
\multicolumn{5}{@{}l}{\textbf{Race (Ref: White)}} \\
\hspace{2mm} Black & \textbf{0.75 (0.67, 0.84)} & 0.92 (0.73, 1.17) & \textbf{0.68 (0.59, 0.79)} & 0.94 (0.77, 1.14) \\
\hspace{2mm} Other & 1.00 (0.83, 1.22) & 0.99 (0.75, 1.31) & 1.06 (0.89, 1.42) & 0.96 (0.73, 1.20) \\
\hspace{2mm} Unknown & \textbf{0.67 (0.44, 0.99)} & 0.88 (0.44, 1.66) & 0.82 (0.47, 1.28) & 1.04 (0.56, 1.97) \\
\addlinespace
\multicolumn{5}{@{}l}{\textbf{Ethnicity (Ref: Non-Hispanic)}} \\
\hspace{2mm} Hispanic/Latino & 0.88 (0.63, 1.17) & 0.73 (0.38, 1.13) & 1.13 (0.82, 1.70) & 1.17 (0.80, 1.94) \\
\hspace{2mm} Unknown & \textbf{0.68 (0.56, 0.83)} & 1.23 (0.80, 2.04) & \textbf{0.71 (0.53, 0.94)} & \textbf{0.65 (0.40, 0.99)} \\
\addlinespace
\multicolumn{5}{@{}l}{\textbf{Insurance (Ref: Commercial)}} \\
\hspace{2mm} Medicare & \textbf{0.53 (0.44, 0.63)} & 1.14 (0.83, 1.54) & \textbf{0.32 (0.26, 0.39)} & 0.86 (0.66, 1.12) \\
\hspace{2mm} Medicaid & \textbf{0.71 (0.57, 0.87)} & 0.90 (0.64, 1.24) & \textbf{0.67 (0.54, 0.83)} & 0.83 (0.61, 1.10) \\
\hspace{2mm} Other & \textbf{0.07 (0.05, 0.09)} & 0.97 (0.54, 1.80) & 0.71 (0.50, 1.02) & 0.96 (0.56, 1.61) \\
\addlinespace
\multicolumn{5}{@{}l}{\textbf{Employment Status (Ref: Full Time)}} \\
\hspace{2mm} Disability & 1.01 (0.85, 1.19) & 1.04 (0.77, 1.44) & \textbf{0.63 (0.51, 0.77)} & 1.23 (0.97, 1.58) \\
\hspace{2mm} Other & \textbf{0.74 (0.61, 0.91)} & 0.90 (0.62, 1.31) & \textbf{0.59 (0.46, 0.75)} & \textbf{0.71 (0.51, 0.98)} \\
\hspace{2mm} Retired & 1.05 (0.88, 1.28) & 1.36 (0.97, 2.22) & 0.98 (0.79, 1.21) & 1.13 (0.89, 1.53) \\
\hspace{2mm} Unknown & \textbf{0.21 (0.17, 0.27)} & 0.71 (0.42, 1.26) & \textbf{0.27 (0.18, 0.41)} & \textbf{0.19 (0.06, 0.46)} \\
\bottomrule
\end{tabular*}

\smallskip
\parbox{\textwidth}{\scriptsize OR: odds ratio; CI: credible interval. Bold values indicate that the 95\% credible interval excludes 1.}
\end{table}

\clearpage
\section*{Appendix D: Simulation study}
\setcounter{table}{0}
\renewcommand{\thetable}{D.\arabic{table}}

This appendix reports the technical details of the simulation studies in
Section 3 of the main text.

\subsection*{D.1 Priors and estimation}

The BSLR is fitted as specified in Sections 2.3 and 2.4 of the main text. The transition-specific coefficients are decomposed as \[ \alpha_j = \alpha + a_j, \qquad \beta_{p,j} = \beta_p + b_{p,j}, \qquad j = 1,\dots,4,\; p = 1,2,3, \] with sum-to-zero constraints $\sum_{j=1}^{4} a_j = 0$ and $\sum_{j=1}^{4} b_{p,j} = 0$, and \[ \alpha \sim N(0, \sigma_\alpha^2), \quad a_j \sim N(0, \sigma_a^2), \quad \beta_p \sim N(0, \sigma_{\beta,p}^2), \quad b_{p,j} \sim N(0, \sigma_{b,p}^2). \] All scale parameters $\sigma_\alpha$, $\sigma_a$, $\sigma_{\beta,p}$ and $\sigma_{b,p}$ are assigned Half-Student-$t$ priors with three degrees of freedom and scale 3.

The ILR fits each transition separately as $\mathrm{logit}(\pi_{ij}) = \alpha_j + X_i^\top \beta_j$, with $\alpha_j \sim N(0, \sigma_\alpha^2)$ and $\beta_{p,j} \sim N(0, \sigma_{\beta,p}^2)$, and the same Half-Student-$t$(3, 3) priors on $\sigma_\alpha$ and $\sigma_{\beta,p}$. The transition-specific deviations and their scale parameters are absent, so that the two models differ only in the partial-pooling structure.

Both models are fitted in Stan using four chains of 6{,}000 iterations, with the first 2{,}000 iterations of each chain discarded as warm-up.

\subsection*{D.2 Performance measures}

Let $\theta$ denote a parameter with true value $\theta_0$, and let
$\widehat\theta^{(r)}$ and $\widehat{V}^{(r)}$ denote its posterior mean and
posterior variance in replicate $r = 1,\dots,R$, with $R = 100$. The measures
reported in Figures 2 and 3 of the main text are
\[
\mathrm{Bias} = \frac{1}{R}\sum_{r=1}^{R}\bigl(\widehat\theta^{(r)} - \theta_0\bigr),
\qquad
\overline{V} = \frac{1}{R}\sum_{r=1}^{R}\widehat{V}^{(r)},
\qquad
\mathrm{MSE} = \frac{1}{R}\sum_{r=1}^{R}\bigl(\widehat\theta^{(r)} - \theta_0\bigr)^2 .
\]
For the covariate effects, panel (b) of Figure 2 reports the bias averaged over the three covariates, denoted $\bar\beta$. The intervals in panel (b) and the bands in panel (d) of Figure 2 and in Figure 3 are computed as the mean across replicates plus or minus 1.96 times its Monte Carlo standard error, the standard deviation across replicates divided by $\sqrt{R}$.

\subsection*{D.3 Data-generating parameters}

Table D.1 reports the true overall effects and transition-specific deviations under the two scenarios, and Table D.2 the implied cumulative and conditional probabilities and the expected number of subjects at risk under the fixed design matrix.

\begin{table}[!htbp]
\spacingset{1}
\centering
\caption{True overall effects and transition-specific deviations in the main
simulation study.}
\label{tab:sim_param}
\begin{tabular}{lrrrrr}
\toprule
 & Overall & \multicolumn{4}{c}{Deviation} \\
\cmidrule(lr){3-6}
Parameter & & $j=1$ & $j=2$ & $j=3$ & $j=4$ \\
\midrule
\multicolumn{6}{l}{\textit{High Probability}} \\
$\alpha$  & $2.806$  & $1.394$  & $0.051$  & $-0.208$ & $-1.237$ \\
$\beta_1$ & $1.164$  & $0.966$  & $-0.174$ & $-0.668$ & $-0.124$ \\
$\beta_2$ & $0.803$  & $0.104$  & $-0.388$ & $0.737$  & $-0.452$ \\
$\beta_3$ & $0.332$  & $-0.545$ & $-0.806$ & $0.873$  & $0.478$  \\
\midrule
\multicolumn{6}{l}{\textit{Low Probability}} \\
$\alpha$  & $-0.835$ & $-0.417$ & $-1.131$ & $1.505$  & $0.043$  \\
$\beta_1$ & $-1.540$ & $1.573$  & $0.286$  & $-0.054$ & $-1.804$ \\
$\beta_2$ & $-0.431$ & $-0.039$ & $0.367$  & $-0.429$ & $0.101$  \\
$\beta_3$ & $-0.364$ & $1.284$  & $0.625$  & $-1.402$ & $-0.508$ \\
\bottomrule
\end{tabular}

\vspace{0.5em}
\parbox{0.85\linewidth}{\footnotesize The transition-specific coefficients are
$\alpha_j = \alpha + a_j$ and $\beta_{p,j} = \beta_p + b_{p,j}$, where the
deviations satisfy $\sum_{j=1}^{4} a_j = 0$ and $\sum_{j=1}^{4} b_{p,j} = 0$.}
\end{table}

\begin{table}[!htbp]
\spacingset{1}
\centering
\caption{Implied probabilities and expected number of subjects at risk in the
main simulation study.}
\label{tab:sim_prob}
\begin{tabular}{lrrrrr}
\toprule
 & Stage 1 & Stage 2 & Stage 3 & Stage 4 & Stage 5 \\
\midrule
\multicolumn{6}{l}{\textit{High Probability}} \\
Cumulative probability  & 1.000 & 0.924 & 0.855 & 0.725 & 0.593 \\
Conditional probability & 0.924 & 0.926 & 0.848 & 0.818 & --    \\
Expected number at risk & 1000  & 924   & 855   & 725   & 593   \\
\midrule
\multicolumn{6}{l}{\textit{Low Probability}} \\
Cumulative probability  & 1.000 & 0.267 & 0.051 & 0.032 & 0.026 \\
Conditional probability & 0.267 & 0.192 & 0.632 & 0.786 & --    \\
Expected number at risk & 1000  & 267   & 51    & 32    & 26    \\
\bottomrule
\end{tabular}

\vspace{0.5em}
\parbox{0.85\linewidth}{\footnotesize The cumulative probability is
$P(Y_i \ge s)$ averaged over subjects. The conditional probability at stage
$s$ is the probability of advancing from stage $s$ to stage $s+1$ among the
subjects who reach stage $s$. The expected number at risk is
$N \times P(Y_i \ge s)$ with $N = 1000$.}
\end{table}

\subsection*{D.4 Duration dependence}

The robustness study of Section 3.3 of the main text is conducted with $N = 4000$ subjects. The data-generating process includes a duration variable $t_i$, the time subject $i$ has spent at a stage, drawn once for each subject and entering the linear predictor of every transition. Its overall effect is negative, so that a longer duration lowers the probability of advancing, and its transition-specific components follow a decreasing gradient across transitions, so that the effect is stronger at later transitions. The duration term is not included in either fitted model; both models are specified as in Section D.1. Table D.3 reports the overall effects and the transition-specific deviations of the intercept and the covariates under the two scenarios, and Table D.4 the implied cumulative probability of reaching each stage.

\begin{table}[!htbp]
\spacingset{1}
\centering
\caption{True overall effects and transition-specific deviations in the
duration-dependence study.}
\label{tab:dur_param}
\begin{tabular}{lrrrrr}
\toprule
 & Overall & \multicolumn{4}{c}{Deviation} \\
\cmidrule(lr){3-6}
Parameter & & $j=1$ & $j=2$ & $j=3$ & $j=4$ \\
\midrule
\multicolumn{6}{l}{\textit{High Probability}} \\
$\alpha$  & $4.042$  & $-0.142$ & $0.474$  & $-0.416$ & $0.084$  \\
$\beta_1$ & $1.936$  & $0.312$  & $1.069$  & $-1.240$ & $-0.142$ \\
$\beta_2$ & $-0.699$ & $0.183$  & $-0.281$ & $3.009$  & $-2.911$ \\
$\beta_3$ & $-1.064$ & $2.357$  & $-0.656$ & $0.309$  & $-2.009$ \\
\midrule
\multicolumn{6}{l}{\textit{Low Probability}} \\
$\alpha$  & $-1.188$ & $1.607$  & $-0.467$ & $-1.181$ & $0.041$  \\
$\beta_1$ & $-0.236$ & $-1.753$ & $2.368$  & $0.903$  & $-1.518$ \\
$\beta_2$ & $-0.922$ & $1.965$  & $1.005$  & $-0.227$ & $-2.743$ \\
$\beta_3$ & $0.288$  & $-2.706$ & $2.331$  & $0.719$  & $-0.345$ \\
\bottomrule
\end{tabular}

\vspace{0.5em}
\parbox{0.85\linewidth}{\footnotesize The transition-specific coefficients are
$\alpha_j = \alpha + a_j$ and $\beta_{p,j} = \beta_p + b_{p,j}$, where the
deviations satisfy $\sum_{j=1}^{4} a_j = 0$ and $\sum_{j=1}^{4} b_{p,j} = 0$.}
\end{table}

\begin{table}[!htbp]
\spacingset{1}
\centering
\caption{Implied cumulative probability of reaching each stage in the
duration-dependence study.}
\label{tab:dur_prob}
\begin{tabular}{lrrrrr}
\toprule
Scenario & Stage 1 & Stage 2 & Stage 3 & Stage 4 & Stage 5 \\
\midrule
High Probability & 1.000 & 0.888 & 0.784 & 0.675 & 0.561 \\
Low Probability  & 1.000 & 0.544 & 0.088 & 0.025 & 0.015 \\
\bottomrule
\end{tabular}

\vspace{0.5em}
\parbox{0.85\linewidth}{\footnotesize The cumulative probability is
$P(Y_i \ge s)$ averaged over subjects.}
\end{table}
\end{document}